\documentclass{article} 

\usepackage{filecontents}  
\usepackage{float}  
\usepackage{cite}  
\usepackage[utf8]{inputenc} 
\usepackage{amsmath}
\usepackage{physics} 
\usepackage{graphicx}
\usepackage{appendix} 
\usepackage[margin=1.6in]{geometry}
\usepackage{booktabs}
\usepackage[affil-it]{authblk} 
\usepackage{etoolbox} 
\usepackage{lmodern}
\usepackage{xr}
\makeatletter
\patchcmd{\@maketitle}{\LARGE \@title}{\fontsize{13.1}{19.2}\selectfont\@title}{}{}
\makeatother

\title{\textbf{
    QED theory of electron beam-induced electronic excitation
    and its effect on sputtering cross sections in 2D crystals
}}
\author[1,2*]{Anthony Yoshimura} 
\author[2]{Michael Lamparski} 
\author[2]{Joel Giedt} 
\author[3]{\\David Lingerfelt} 
\author[3]{Jacek Jakowski} 
\author[3]{Panchapakesan Ganesh} 
\author[4]{Tao Yu}
\author[3]{\\Bobby Sumpter} 
\author[2,5]{Vincent Meunier} 
\affil[1]{Lawrence Livermore National Laboratory, Livermore, CA 94550, USA} 
\affil[2]{Department of Physics, Applied Physics, and Astronomy,
Rensselaer Polytechnic Institute, Troy, NY 12180, USA}
\affil[3]{Center for Nanophase Material Sciences, Oak Ridge National
Laboratory, Oak Ridge, TN 37831, USA} 
\affil[4]{Department of Chemistry, University of North Dakota, Grand Forks, ND
58202, USA} 
\affil[5]{Department of Materials Science and Engineering, Rensselaer
Polytechnic Institute, Troy, NY 12180, USA} 
\affil[*]{Correspondence to be addressed to yoshimura4@llnl.gov}

\date{}

\begin{document}

\maketitle

\begin{abstract}
  Many computational models have been developed to predict the rates of atomic
  displacements in two-dimensional (2D) materials under electron beam
  irradiation.
  However, these models often drastically underestimate the displacement rates
  in 2D insulators, in which beam-induced electronic excitations can reduce the
  binding energies of the irradiated atoms.
  This bond softening leads to a qualitative disagreement between theory and
  experiment, in that substantial sputtering is experimentally observed at beam
  energies deemed far to small to drive atomic dislocation by many current
  models.
  To address these theoretical shortcomings, this paper develops a
  first-principles method to calculate the probability of beam-induced
  electronic excitations by coupling quantum electrodynamics (QED) scattering
  amplitudes to density functional theory (DFT) single-particle orbitals.
  The presented theory then explicitly considers the effect of these electronic
  excitations on the sputtering cross section.
  Applying this method to 2D hexagonal BN and MoS$_2$ significantly increases
  their calculated sputtering cross sections and correctly yields appreciable
  sputtering rates at beam energies previously predicted to leave the crystals
  intact.
  The proposed QED-DFT approach can be easily extended to describe a rich
  variety of beam-driven phenomena in any crystalline material.
\end{abstract}
\pagebreak
    
\section{Introduction} 
\label{sec:introduction}

The holy grail of materials engineering is atomic scale control of the
material structure.
Towards this aim, electron irradiation by transmission electron microscopy
(TEM) can be an effective means of structural manipulation with spatial control
\cite{Banhart1999, Egerton2013, Zhao2017, Susi2019}.
Structural changes under electron irradiation can arise via atomic displacement
in which an incident electron nudges a material atom from its initial site.  We
call this atom the primary knock-on atom (PKA).  Two-dimensional (2D) crystals
provide an excellent platform to measure the rates of these PKA displacements.
When electron irradiation is normal to a 2D crystal's surface, a displacement
likely propels the PKA away from the crystal.
As such, these displaced atoms are often ejected from the crystal in a process
called \textit{sputtering}.  Sputtering events leave behind vacancies, which
can then be counted using TEM.
Counting the number of vacancies for a given dosage and beam energy allows one
to experimentally determine the sputtering cross section of that crystal.

Sputtering occurs when the energy transferred to the PKA is greater than
the PKA's displacement threshold $E_d$.
This means that a displacement is possible only if the kinetic energy of the
beam electron exceeds some critical energy $\epsilon_c$.
Many computational models have been proposed to predict both $E_d$ and
$\epsilon_c$ to calculate electron beam-induced sputtering rates in 2D crystals
\cite{Meyer2012, Susi2016, Yoshimura2018, Susi2019}.
However, the vast majority of current
methods focus solely on interactions between the beam electrons and material
nuclei, neglecting any coupling with the material’s electrons.
Thus, while present-day models give reasonable predictions for conductors
\cite{Meyer2012}, where electronic relaxation is rapid, they often vastly
underestimate the atomic displacement rates in insulators.
For example, the critical energy for sputtering boron or nitrogen from
hexagonal boron nitride (hBN) is predicted to be 80 keV \cite{Kotakoski2010}.
However, sputtering has been observed in hBN under 30 keV irradiation
\cite{Cretu2015}.
Furthermore, selenium sputters from WSe$_2$ and MoSe$_2$ under irradiation
energies of 60 and 80 keV, respectively.  These energies are almost 150 keV
below their predicted critical energies \cite{Lin2015, Lehnert2017}.
Lastly, while the calculated critical energy for sulfur sputtering in MoS$_2$
is about 90 keV, sulfur has been shown to sputter under 20 keV beams
\cite{Kretschmer2020}.
Discrepancies like these suggest that the displacement thresholds in insulating
crystals are much smaller than what is predicted by ground-state theory.
Lehnert et al.  have proposed that the consideration of inelastic scattering,
i.e., beam-induced electronic excitation, can lead to such a reduction in the
displacement threshold \cite{Lehnert2017}.
This would increase the sputtering cross section for all beam energies and
enable sputtering for energies well-below the ground state $\epsilon_c$.

To account for these effects, we combine quantum electrodynamics (QED) and
density functional theory (DFT) to derive the probability of beam-induced
electronic excitation in 2D insulating crystals.
The basic idea is as follows: DFT can provide effective single-particle states
that can be decomposed into a plane-wave basis \cite{Hohenberg1964, Kohn1965,
Kresse1996a}, while QED is well-equipped to describe how each plane-wave
evolves in time through interactions with an electromagnetic field
\cite{Lourenco-Martins2021, Peskin1995, Lancaster2014}.
Thus, a plane-wave decomposition of the Kohn-Sham orbitals can allow for a
component-by-component treatment of the interactions between the beam and
material electrons.
This generalized QED-DFT approach enables, for the first time, a
first-principles description of any beam-matter interaction process.
The only limitations of this method are the order to which the time-evolution
operator is expanded and the sophistication of the theory used to determine the
material's electronic structure.
Additionally, while DFT is used here, our method is compatible with any
first-principles formalism that can produce single-particle eigenstates and
eigenvalues for a given material.

This paper is divided into three sections.
First, we describe the key physical processes that dictate the rate of
beam-induced sputtering in a 2D insulating crystal.
Second, we derive the probability of beam-induced electronic excitation in
these materials as a function of beam energy.
Finally, we show how this excitation probability can be used to predict
sputtering cross sections in hBN and MoS$_2$ that quantitatively agree with
experiment.


\section{Three interactions and three rates}
\label{sec:bigPicture}

The majority of present-day beam-damage models focus solely on the interaction
between the beam electron and target nucleus.
These models are thus centered on one process: energy transfer from the beam
electron to the nucleus.
For this, one defines a differential cross section $d\sigma/dE(\epsilon_b, E)$
providing the distribution of energy transfers $E$ for a given beam energy
$\epsilon_b$.
The McKinley and Feshbach differential cross section can adequately describe
$d\sigma/dE$
for light nuclei (Z $<$ 20) \cite{Mott1929, McKinley1948, Oen1973,
Egerton2010}.  Setting $\hbar = c = 1$,

\begin{equation}
  \frac{d\sigma}{dE}
  =
  \pi\left(\frac{Z\alpha}{|\mathbf{p}_b|\beta}\right)^2
  \left[
    \frac{E_\text{max}}{E^2}
    - \beta(\pi Z\alpha+\beta)\frac{1}{E}
    + \pi Z\alpha\beta\sqrt{\frac{E_\text{max}}{E^3}}
  \right],
  \label{eq:MF}
\end{equation}
where $\mathbf{p}_b$ is the momentum of the beam electron, $\beta$ is its
velocity, $Z$ is the atomic number of the target nucleus, and $\alpha$ is the
fine structure constant.

One can then calculate the displacement cross section by integrating
$d\sigma/dE$ over all $E$ large enough to cause a displacement, so that

\begin{equation}
  \sigma_0(\epsilon_b, {E}_d)
  =
  \theta(E_\text{max}(\epsilon_b) - E_d)
  \int_{E_d}^{E_\text{max}(\epsilon_b)}
  \frac{d\sigma}{dE}(\epsilon_b, E)
  dE,
  \label{eq:basicSigma}
\end{equation}
where $E_\text{max}(\epsilon_b)$
is the maximum possible energy transfer for a given beam energy, i.e, the
energy transfer resulting from a direct collision, and the step function
$\theta$ enforces that the cross section is zero when $E_\text{max} < E_d$.
We will leave the step function implicit going forward for the sake of
compactness.
The critical energy $\epsilon_c$ is then the beam energy for which
$E_\text{max}=E_d$.
Therefore, the observation of sputtering at energies well below $\epsilon_c$ is
completely at odds with equation (\ref{eq:basicSigma}).
With this in mind, a fair amount of work has been done to treat deviations from
equation (\ref{eq:basicSigma}).
Notably, several studies have explored the effects of temperature on
displacement cross sections 
\cite{Meyer2012,Susi2016,Yoshimura2018}.
This consideration involves calculating the degree to which the pre-collision
thermal motion of the nucleus increases the cross section.
However, these techniques essentially amount to smearing the beam energy
dependence of the cross section, so that the cross section only strays
significantly from equation (\ref{eq:basicSigma}) for beam energies very close
to $\epsilon_c$.
Thus, temperature-induced increases in the cross section cannot account for the
disparities between the equation (\ref{eq:basicSigma}) model and
experiment.
This necessitates the consideration of additional phenomena that can reduce
$E_d$.

\begin{figure}
  \centering
  \includegraphics[width=.7\textwidth]{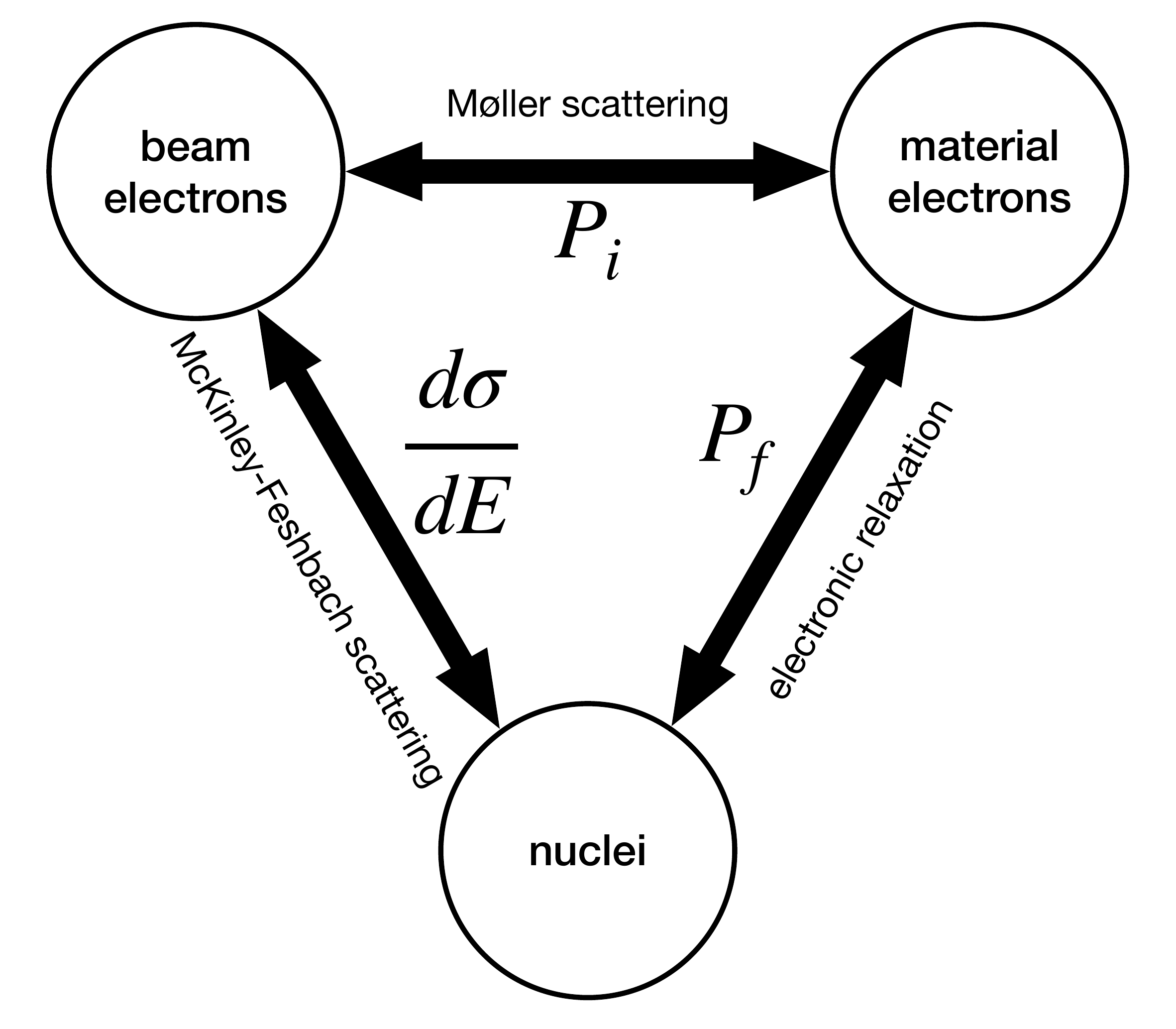}
  \caption{
    Three interactions between three pairs of particles involved in electron
    beam-induced sputtering.
    In insulators, the probability of material electronic excitation ($P_i$)
    and the probability that those excitations are substantially long-lived
    ($P_f$) can be large.
    Therefore, the effect of exciting the material electrons on the sputtering
    rate should not be ignored in these materials.
  }
  \label{fig:triad}
\end{figure}

To address the limitations of equation (\ref{eq:basicSigma}), this work
introduces a third party: the material's electrons.
Doing so brings two new interactions into play: one between the beam and
material electrons and another between material electrons and nuclei.
This yields a total of three interactions between the three pairs of particles
(figure \ref{fig:triad}).
Therefore, the rate of beam-induced sputtering hinges on the rates of three
processes mediated by these interactions.

\begin{enumerate}
  \item Beam and material electrons:
    a beam electron can excite some number $n_i$ ground state electrons to the
    conduction band ($i$ denotes the \textit{initial} interaction with the
    beam).
    The probability of this event for a given beam energy $\epsilon_b$ is
    $P_i(\epsilon_b, n_i)$.
  \item Material electrons and nuclei:
    Some number $n_f$ beam-induced excitations can survive long enough for the
    target atom to leave its original site ($f$ denotes the \textit{final}
    system at the completion of sputtering).
    This depends on the nuclear kinetic energy $E$ and the excitation lifetime
    $\tau$.
    The probability that $n_f$ of the $n_i$ excitations survive is $P_f(E,
    \tau, n_i, n_f)$.
  \item Beam electrons and material nuclei: the energy transferred to a
    material nucleus by the beam electron can exceed the PKA's displacement
    threshold $E_d(n_f)$, which depends on the number of surviving excitations
    $n_f$.  We define $\{{E}_d\}$ as the set of all displacement thresholds for
    all possible $n_f$.  Sputtering occurs when $E>E_d(n_f)$.  The differential
    cross section for an energy transfer $E$ from the beam electron to
    material nucleus is $d\sigma/dE(\epsilon_b, E)$.
\end{enumerate}

\noindent
The sputtering cross section can then be calculated by coupling $d\sigma/dE$ to
$P_i$ and $P_f$ for all possible $n_i$ and $n_f$.
With the terms defined above, this excitation-sensitive sputtering
cross section can be written as

\begin{equation}
  \begin{aligned}
    \sigma(\epsilon_b, \{{E}_d\}, \tau)
    &=
    \sum_{n_i=0}^\infty
    P_i(\epsilon_b, n_i)
    \sum_{n_f=0}^{n_i}
    \int_{E_d(n_f)}^{E_\text{max}(\epsilon_b)}
    P_f(E, \tau, n_i, n_f)
    \frac{d\sigma}{dE}(\epsilon_b, E)
    dE.
  \end{aligned}
  \label{eq:totSigma}
\end{equation}
%
If $P_i$ and $P_f$ are non-negligible when $n_i$ and $n_f$ are nonzero, and
$E_d$ depends strongly on $n_f$, then interactions with the material electrons
must be considered.
We will later show that this makes $\sigma$ in equation (\ref{eq:totSigma})
larger than $\sigma_0$ in equation (\ref{eq:basicSigma}) for all beam energies,
most prominently when $\epsilon_b < \epsilon_c$.

The remainder of this paper focuses on the derivations of $P_i$, $P_f$, and
$\{E_d\}$.
Section \ref{sec:probability} describes how to combine QED with DFT to obtain
$P_i$.
Section \ref{sec:sputtering} then considers the evolution of the excited states
during the sputtering process to derive $\{E_d\}$ and $P_f$. It then demonstrates
how our formalism significantly improves the prediction of sputtering rates in
hBN and MoS$_2$.

\section{Probability of beam-induced excitation} 
\label{sec:probability}

For a crystal in its ground state, an occupied electron
energy eigenstate has zero overlap with any unoccupied state.
However, the collision of a beam electron can give an occupied state a
momentum boost that breaks this orthogonality.
Thus, the boosted ground state has a nonzero probability of being measured in
an excited state.
We can use this idea to derive $P_i(\epsilon_b, n_i)$, the probability that a
beam electron with kinetic energy $\epsilon_b$ excites exactly $n_i$ material
electrons.
The derivation can be broken down into four steps:
(i)
determine the amplitude for a free electron to scatter from one momentum
eigenstate into another after collision with another free electron;
(ii)
generalize the formalism to obtain the amplitude for scattering from one wave
packet into another by summing over the amplitudes for each momentum component
of one wave packet to scatter into each momentum component of the other;
(iii)
decompose a pair of occupied and unoccupied crystal states into a momentum
basis and plug them in as incoming and outgoing wave packets respectively, then
square the amplitude to obtain the corresponding excitation probability for a
particular transition;
(iv)
Compute the sum of all transition probabilities and use combinatorics to
determine $P_i(\epsilon_b, n_i)$.
The following subsections address each step (i-iv) in detail.

\subsection{Scattering of free electrons} 
\label{sec:ee}

\begin{figure}
  \centering
  \includegraphics[width=.9\textwidth]{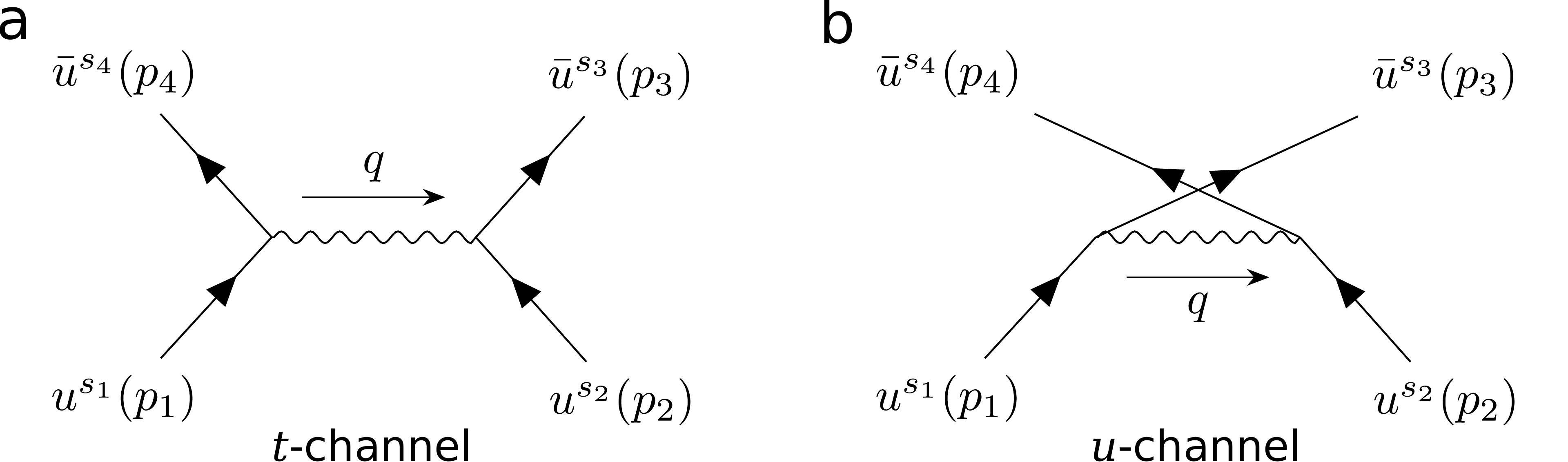}
  \caption{
    The lowest order electron-electron scattering perturbation includes two
    Feynman diagrams called the (a) $t$-channel and (b) $u$-channel.
    The incoming and outgoing electron states are represented by Dirac spinors
    $u^s(p)$ and $\bar{u}^s(p)$ respectively, where $p$ and $s$ are the electron's
    4-momentum and spin index respectively.
    Subscripts 1 and 2 label components of the initial beam and material states
    respectively.
    The virtual photon 4-momentum $q$ is the momentum transfer between the
    electrons.
  }
\label{fig:tu}
\end{figure}

We first derive the scattering amplitude for momentum transfer between two free
electrons via M{\o}ller scattering \cite{Moller1932, Kragh1992, Roqu1992}.
With this, the interaction between actual material states can be written as a
linear combination of these free-particle interactions.
Going forward, we label the 4-momenta of the incoming electrons as $p_1$ and
$p_2$, while the outgoing electrons have momenta $p_3$ and $p_4$.
We also choose to make $p_1$ and $p_2$ components of the initial beam and
material states respectively.
The 4-momentum of the $n$th electron can be written as
$p_n = (\epsilon_n, p_n^x, p_n^y, p_n^z) = (\epsilon_n, \mathbf{p}_n)$,
where $\epsilon_n$ is the particle's energy and $\mathbf{p}_n$ is its
3-momentum.
Dot products between 4-vectors are then taken over Minkowski space, so that
$p_n\cdot p_m
=
g_{\mu\nu}p_n^\mu p_m^\nu
=
\epsilon_n\epsilon_m -
\mathbf{p}_n\cdot\mathbf{p}_m$,
where
$\mathbf{p}_n\cdot\mathbf{p}_m
=
p_n^xp_m^x + p_n^yp_m^y + p_n^zp_n^z$.   

To lowest order, the amplitude for free electron scattering can be represented
by two tree-level diagrams, which we call the $t$- and $u$-channels (figure
\ref{fig:tu}).
Using Feynman's rules \cite{Peskin1995, Lancaster2014}, we can write these
diagrams in terms of Dirac spinors, yielding the invariant matrix element

\begin{equation} 
  \label{eq:M} 
  \begin{aligned} 
    \mathcal{M}(p_4p_3\leftarrow
    p_2p_1) 
    &=
    \frac{e^2}{2} \sum_{s_1}\sum_{s_2}\sum_{s_3}\sum_{s_4} 
    \\& \bigg[
      \bar{u}^{s_4}\left(p_4\right)\gamma^{\mu}u^{s_1}\left(p_1\right)
      \left(\frac{1}{p_3 - p_2}\right)^2
      \bar{u}^{s_3}\left(p_3\right)\gamma_{\mu}u^{s_2}\left(p_2\right)
      \\&+ 
      \bar{u}^{s_3}\left(p_3\right)\gamma^{\mu}u^{s_1}\left(p_1\right)
      \left(\frac{1}{p_4 - p_2}\right)^2
      \bar{u}^{s_4}\left(p_4\right)\gamma_{\mu}u^{s_2}\left(p_2\right)
    \bigg],
  \end{aligned} 
\end{equation}
where $s_n = 1\text{ or }2$ denotes the spin of the $n$th electron, $u^s(p)$
is a Dirac spinor, and $\bar{u}^s(p)$ is its conjugate (section \ref{app:M}).
The factor of 1/2 before the summation arises from the assumption that the
incoming states are spin unpolarized.

The first term in brackets is the $t$-channel describing momentum transfer $p_3
- p_2$ and the second is the $u$-channel describing momentum transfer $p_4 -
p_2$.
Because the DFT cutoff energy is much smaller than the beam energy, it is
always the case that $|\mathbf{p}_2|$ is much smaller than $|\mathbf{p}_1|$.
Furthermore, we need only consider outgoing momenta for which the kinetic
energy associated with either $|\mathbf{p}_3|$ or $|\mathbf{p}_4|$ falls within
the DFT cutoff energy.
In these cases, the magnitude of one outgoing momentum is similar to
$|\mathbf{p}_2|$, while that of the other is much greater.
This means that one channel's momentum transfer is always much larger than the
other's.
As the momentum transfers reside in the denominators of either channel in
equation (\ref{eq:M}), it follows that one channel always contributes much more
to $\mathcal{M}$ than the other.
Thus, when the $t$-channel is significant, the $u$-channel is negligible, and
vice versa.
Additionally, when integrating over all possible outgoing momenta, the
contribution of the $t$-channel is equal to that of the $u$-channel.
Taking advantage of this along with the indistinguishably of the electrons, we
calculate only the $t$-channel and multiply the resulting amplitude by 2
instead of calculating both channels and adding them.
We can then define the 4-momentum transfer between the electrons as that of the
$t$-channel: $q\equiv p_3 - p_2$.
Because the $t$-channel has a $q^2$ in the denominator, the resulting
scattering probability is proportional to $q^{-4}$.
This makes large momentum transfers statistically irrelevant, allowing us to
only consider momentum transfers inside the first Brillouin zone (BZ).
The evaluation of the $t$-channel in terms of the
components of the electrons' 4-momenta is straightforward, though cumbersome,
and is described in section \ref{app:M}.

We can use the resulting $\mathcal{M}$ to obtain the free electron scattering
amplitude

\begin{equation} 
  \label{eq:T} 
  \bra{p_4 p_3}\hat{T}\ket{p_2p_1} 
  = 
  (2\pi)^4\delta(p_1 + p_2 - p_3 - p_4) 
  \mathcal{M}(p_4p_3\leftarrow p_2p_1),
\end{equation}
where $\hat{T}$ is the scattering operator \cite{Peskin1995, Lancaster2014}.
This gives the amplitude for two free electrons with momenta $p_1$ and $p_2$ to
scatter into $p_3$ and $p_4$.
Equation (\ref{eq:T}) is used to derive the scattering amplitude between two
arbitrary wave packets in the next subsection.

\subsection{Scattering of wave packets} 
\label{sec:wavepackets}


The free particle scattering amplitude in equation (\ref{eq:T}) can be used to
determine the amplitude for the scattering of two arbitrary electron states
$\phi_1$ and $\phi_2$ into $\phi_3$ and $\phi_4$.
This is obtained by sandwiching the scattering operator between the initial and
final 2-particle states, i.e.,

\begin{equation} 
  \label{eq:amp} 
  \begin{aligned}
    &\bra{\phi_4\phi_3} \hat{T} \ket{\phi_2\phi_1}
    = 
    \int \frac{d^3p_4d^3p_3d^3p_2d^3p_1} {(2\pi)^{12}
    16\epsilon_4\epsilon_3\epsilon_2\epsilon_1}
    \bra{\phi_4\phi_3}\ket{p_4p_3} \bra{p_4p_3}\hat{T}\ket{p_2p_1}
    \bra{p_2p_1}\ket{\phi_2\phi_1}.
\end{aligned} 
\end{equation}
On the right side, we have inserted two resolutions of the identity given in
equation (\ref{eq:identity}). Inserting equation (\ref{eq:T}) into the
integrand, the amplitude can be written in terms of the invariant matrix
element $\mathcal{M}$, becoming

\begin{equation} 
  \label{eq:ampPlugInM} 
  \begin{aligned}
    &\int\frac{d^3p_4d^3p_3d^3p_2d^3p_1} {(2\pi)^{12}
    16\epsilon_4\epsilon_3\epsilon_2\epsilon_1} \bra{\phi_4}\ket{p_4}
    \bra{\phi_3}\ket{p_3} \bra{p_2}\ket{\phi_2} \bra{p_1}\ket{\phi_1}
    \\&\qquad\times 
    \mathcal{M}(p_4p_3\leftarrow p_2p_1) 
    (2\pi)^4\delta(p_1 + p_2 - p_3 - p_4).  
\end{aligned} 
\end{equation}
Using the delta function to integrate over $\mathbf{p}_4$ and $p_3^z$ yields

\begin{equation} 
  \label{eq:ampUseDelta} 
  \begin{aligned} 
    \int\frac{d^2p_3^\perp
    d^3p_2d^3p_1} {(2\pi)^{8} 16\epsilon_2\epsilon_1}
    \frac{\mathcal{M}(p_4p_3\leftarrow p_2p_1)}
    {\left|p_3^z\epsilon_4 - p_4^z\epsilon_3\right|} \bra{\phi_4}\ket{p_4}
    \bra{\phi_3}\ket{p_3} \bra{p_2}\ket{\phi_2} \bra{p_1}\ket{\phi_1},
  \end{aligned} 
\end{equation}
where it is understood that $p_4$ and $p_3$ satisfy $p_1 + p_2 = p_3 + p_4$
(section \ref{app:p3z}).
The normalization of 4-momentum states $\ket{p_n}$ in terms of
3-momentum states $\ket{\mathbf{p}_n}$ as defined in equation
(\ref{eq:normalization}) allows us to rewrite the expression as

\begin{equation} 
  \label{eq:amp3Mom} 
  \begin{aligned} 
    \int\frac{d^2p_3^\perp
    d^3p_2d^3p_1}{(2\pi)^{8}}
    \sqrt{\frac{\epsilon_4\epsilon_3}{\epsilon_2\epsilon_1}}
    \frac{\mathcal{M}(p_4p_3\leftarrow p_2p_1)} {4\left|p_3^z\epsilon_4 -
    p_4^z\epsilon_3\right|} \bra{\phi_4}\ket{\mathbf{p}_4}
    \bra{\phi_3}\ket{\mathbf{p}_3} \bra{\mathbf{p}_2}\ket{\phi_2}
    \bra{\mathbf{p}_1}\ket{\phi_1}.  
  \end{aligned} 
\end{equation}
We can then discretize the momenta by replacing $d^3p_i/(2\pi)^3$ with $V^{-1}$
and $d^2p_i^\perp/(2\pi)^2$ with $A^{-1}$, where $V$ and $A$ are the volume and
cross sectional area of the simulated crystal, i.e., the volume and
cross sectional area of the unit cell times the number of k-points used to
sample the BZ.
With this, the amplitude for electron states $\ket{\phi_1}$ and $\ket{\phi_2}$
to scatter into $\ket{\phi_3}$ and $\ket{\phi_4}$ takes the form

\begin{equation} 
  \label{eq:ampDisc} 
  \begin{aligned} 
    \frac{1}{AV^2}
    \sum_{\mathbf{p}_3^\perp} \sum_{\mathbf{p}_2} \sum_{\mathbf{p}_1}
    \sqrt{\frac{\epsilon_4\epsilon_3}{\epsilon_2\epsilon_1}}
    \frac{\mathcal{M}(p_4p_3\leftarrow p_2p_1)} {4\left|p_3^z\epsilon_4 -
    p_4^z\epsilon_3\right|} \bra{\phi_4}\ket{\mathbf{p}_4}
    \bra{\phi_3}\ket{\mathbf{p}_3} \bra{\mathbf{p}_2}\ket{\phi_2}
    \bra{\mathbf{p}_1}\ket{\phi_1}.  
  \end{aligned} 
\end{equation}
In the next subsection, we replace $\ket{\phi_{1\dots 4}}$ with states relevant to
electron beam-induced excitation.

\subsection{Probability of a crystal excitation} 
\label{sec:crystal}

We now consider the specific case of beam-induced excitations to determine the
form of the four electron states in equation (\ref{eq:ampDisc}).  We assign
$\ket{\phi_1}$ and $\ket{\phi_4}$ to the initial and final beam states
$\ket{\mathbf{p}_b}$ and $\ket{\mathbf{p}_b'}$ respectively.  States $\ket{\phi_2}$ and
$\ket{\phi_3}$ are then the ground and excited crystal states
$\ket{n\mathbf{k}}$ and $\ket{n'\mathbf{k'}}$ respectively, where $n$ and $n'$
are band indices and $\mathbf{k}$ and $\mathbf{k'}$ are k-points.
Substituting these specific states into expression (\ref{eq:ampDisc}), the
amplitude for exciting $\ket{n\mathbf{k}}$ to $\ket{n'\mathbf{k'}}$ becomes

\begin{equation} 
  \label{eq:ampCrystal} 
  \begin{aligned}
    \bra{\mathbf{p}_b', n'\mathbf{k'}} \hat{T} \ket{n\mathbf{k},\mathbf{p}_b}
    &=
    \frac{1}{AV^2}
    \sum_{\mathbf{p}_3^\perp} \sum_{\mathbf{p}_2} \sum_{\mathbf{p}_1}
    \sqrt{\frac{\epsilon_4\epsilon_3}{\epsilon_2\epsilon_1}}
    \frac{\mathcal{M}(p_4p_3\leftarrow p_2p_1)}
    {4\left|p_3^z\epsilon_4 - p_4^z\epsilon_3\right|}
    \\&\times
    \bra{\mathbf{p'}_b}\ket{\mathbf{p}_4}
    \bra{n'\mathbf{k'}}\ket{\mathbf{p}_3}
    \bra{\mathbf{p}_2}\ket{n\mathbf{k}}
    \bra{\mathbf{p}_1}\ket{\mathbf{p}_b}.
  \end{aligned}
\end{equation}
The values of $\epsilon_{1\dots4}$ need to be clarified before moving forward.
The zeroth components of the initial and final beam momenta obey the free
particle dispersion relations, so 
$\epsilon_1 = \sqrt{|\mathbf{p}_1|^2 + m^2}$
and
$\epsilon_4 = \sqrt{|\mathbf{p}_4|^2 + m^2}$.
The beam energy that appears in equation (\ref{eq:totSigma}) is then
defined as the beam electron's total energy minus its rest mass:
$\epsilon_b\equiv\epsilon_1 - m$.
Meanwhile, the momentum of the crystal states can be treated
nonrelativistically.
Thus, the zeroth components of the crystal state momenta
are the energy eigenvalues of the crystal state plus the electron rest mass,
i.e.,
$\epsilon_2=\epsilon_{n\mathbf{k}} + m$
and
$\epsilon_3=\epsilon_{n'\mathbf{k'}} + m$.
For the remainder of this derivation, we continue to leave our expressions
in terms of $\epsilon_{1\dots4}$ for compactness.

Sputtering from a 2D crystal often requires that the beam electron is
backscattered or nearly backscattered, in which case, its final trajectory
after collision with the nucleus is nearly antiparallel to its initial
trajectory and perpendicular to the crystal surface.
Given that many 2D materials (including hBN and MoS$_2$) possess inversion
and/or reflection symmetry about the crystal plane, we assume that the
likelihood of excitation before and after the collision are about equal.
In light of this, we calculate the excitation probability during a sputtering event
assuming the beam electron's trajectory is not altered by its collision with
the nucleus.
That is, we impose that $\mathbf{p}_1 = |\mathbf{p}_1|\hat{\mathbf{z}}$ until
an electronic excitation is induced.

We can now evaluate the bra-ket products in equation (\ref{eq:ampCrystal}).
The initial beam state is highly localized on $\mathbf{p}_b$, meaning that

\begin{equation} 
  \label{eq:phi_1}
    |\bra{\mathbf{p}_1}\ket{\mathbf{p}_b}|^2
    =
    V\delta_{\mathbf{p}_1,\mathbf{p}_b}
    \quad\Rightarrow\quad
    \bra{\mathbf{p}_1}\ket{\mathbf{p}_b}
    =
    V^{1/2}\delta_{\mathbf{p}_1,\mathbf{p}_b}.
\end{equation}
Meanwhile, the ground and excited crystal states can be expanded into a
plane-wave basis, so that

\begin{equation} 
  \begin{aligned}
    &\ket{n\mathbf{k}}
    =
    V^{-1/2}\sum_{\mathbf{G}}
    C_{\mathbf{G+k}}^n \ket{\mathbf{G+k}}
    \\& \ket{n'\mathbf{k'}}
    =
    V^{-1/2}\sum_{\mathbf{G}} C_{\mathbf{G+k'}}^{n'} \ket{\mathbf{G+k'}},
  \end{aligned}
\end{equation}
where each $\mathbf{G}$ is a reciprocal lattice vector.
By re-expressing $\mathbf{p}_2$ and $\mathbf{p}_3$ as $\mathbf{G}_2+\mathbf{k}_2$ and
$\mathbf{G}_3 + \mathbf{k}_3$ respectively, we find

\begin{equation} 
  \label{eq:phi_2} 
  \begin{aligned}
    &\bra{\mathbf{p}_2}\ket{n\mathbf{k}}
    =
    \bra{\mathbf{G}_2 + \mathbf{k}_2}
    \left(
        V^{-1/2}\sum_{\mathbf{G}}
        C_{\mathbf{G+k}}^n \ket{\mathbf{G+k}}
    \right)
    =
    V^{1/2}C^n_{\mathbf{G}_2+\mathbf{k}}
    \delta_{\mathbf{k}_2,\mathbf{k}}.
    \\&
    \bra{\mathbf{p}_3}\ket{n'\mathbf{k'}}
    =
    \bra{\mathbf{G}_3 + \mathbf{k}_3}
    \left(
        V^{-1/2}\sum_{\mathbf{G}}
        C_{\mathbf{G+k'}}^{n'} \ket{\mathbf{G+k'}}
    \right)
    =
    V^{1/2}C^{n'}_{\mathbf{G}_3+\mathbf{k'}}
    \delta_{\mathbf{k}_3,\mathbf{k'}}.
  \end{aligned}
\end{equation}
Lastly, we do not care where the outgoing scattered electron ends up, so we
wish for $\ket{\mathbf{p}_b'}$ to satisfy

\begin{equation} 
  \label{eq:phi_4}
    |\bra{\mathbf{p}_4}\ket{\mathbf{p}_b'}|^2
    =
    V \quad\Rightarrow\quad \bra{\mathbf{p}_4}\ket{\mathbf{p}_b'}
    =
    V^{1/2}.
\end{equation}
The excitation amplitude is then obtained by plugging in the bra-ket products from
equations (\ref{eq:phi_1}), (\ref{eq:phi_2}), and (\ref{eq:phi_4}) into
equation (\ref{eq:ampCrystal}). This gives us the excitation amplitude

\begin{equation} 
  \begin{aligned}
      \bra{\mathbf{p}_b', n'\mathbf{k'}} \hat{T} \ket{n\mathbf{k},\mathbf{p}_b}
      =
      \frac{1}{A}
      \sum_{\mathbf{G}_3^\perp} \sum_{\mathbf{G}_2}
      \sqrt{\frac{ \epsilon_4\epsilon_3 }{ \epsilon_1\epsilon_2 }}
      \frac{\mathcal{M}(p_4p_3\leftarrow p_2p_1)}
      {4\left|p_3^z\epsilon_4 - p_4^z\epsilon_3\right|}
      C_{\mathbf{G}_3+\mathbf{k'}}^{n'\ast} C_{\mathbf{G}_2+\mathbf{k}}^{n},
  \end{aligned}
  \label{eq:ampCode} 
\end{equation}
where it is understood that $p_2 = (m + \epsilon_{n\mathbf{k}},
\mathbf{G}_2 + \mathbf{k})$ and
$p_3 = (m + \epsilon_{n'\mathbf{k'}}, \mathbf{G}_3 + \mathbf{k'})$,
%
and $p_4$ and $p_3^z$ satisfy $p_1 + p_2 = p_4 + p_3$.
Squaring this amplitude yields the probability of a single electronic
excitation from the valence band state $\ket{n\mathbf{k}}$ to the conduction
band state $\ket{n'\mathbf{k'}}$ for a given beam energy $\epsilon_b$, that is,

\begin{equation}
  P(n'\mathbf{k'}\leftarrow n\mathbf{k}|\epsilon_b)
  =
  \left|
  \bra{\mathbf{p}_b', n'\mathbf{k'}} \hat{T} \ket{n\mathbf{k},\mathbf{p}_b}
  \right|^2.
  \label{eq:prob}
\end{equation}
%

\subsection{Probability of $n_i$ excitations}
\label{sec:Pi}

We are finally ready to derive $P_i(\epsilon_b, n_i)$, the probability that a
beam electron excites a particular number of electrons $n_i$.  First, we define
the sum of all transition probabilities

\begin{equation}
  S(\epsilon_b)
  \equiv
  \sum_\mathbf{k} \sum_\mathbf{k'} \sum_n \sum_{n'}
  P(n'\mathbf{k}'\leftarrow n\mathbf{k}|\epsilon_b)
  \equiv
  \sum_j^N P_j(\epsilon_b),
  \label{eq:S}
\end{equation}
where $\mathbf{k}$ and $\mathbf{k'}$ run over all k-points, $n$ over
the valence bands, and $n'$ over the conduction bands (those possessing states
with energy between the Fermi level and the work function).
The index $j$ in the right-most expression labels the possible single-particle
excitations (e.g., $j=n'\mathbf{k'}\leftarrow n\mathbf{k}$).

Combinatorics tells us that the probability of exciting exactly one excitation
is

\begin{equation}
  \begin{aligned}
    P_i(n_i=1)
    &=
    \sum_{j_1} P_{j_1}
    \prod_{j_2\neq j_1} (1 - P_{j_2})
    \\&=
    \sum_{j_1} P_{j_1}
    \left(
      1 - \sum_{j_2\neq j_1} P_{j_2}
      + \frac{1}{2} \sum_{j_3\neq j_1,j_2} \sum_{j_2\neq j_1} P_{j_2} P_{j_3}
      - \dots
    \right).
  \label{eq:Pini1}
  \end{aligned}
\end{equation}
In the large-crystal limit, the number of states, and thus the number of
transitions, is large so that the summations over $j_i$ in equation
(\ref{eq:Pini1}) are approximately equal to one another.
That is,

\begin{equation}
  \sum_{j_2\neq j_1}P_{j_2}
  \sim
  \sum_{j_2}P_{j_2}
  =
  S.
\end{equation}
In this limit, the probability of exactly one beam-induced excitation can be
written as

\begin{equation}
  \begin{aligned}
  P_i(n_i=1)
  \sim
  \sum_j P_j
  \left( 1 - S + \frac{1}{2}S^2 - \dots \right)
  =
  e^{-S}\sum_j P_j
  =
  Se^{-S}.
\end{aligned}
\label{eq:Pi(1)}
\end{equation}
In the same way, the probability of two excitations is

\begin{equation}
\begin{aligned}
  P_i(n_i=2)
  &=
  \frac{1}{2} \sum_{j_1} \sum_{j_2\neq j_1} P_{j_1} P_{j_2}
  \prod_{j_3\neq j_1,j_2} (1 - P_{j_3})
  \sim
  \frac{S^2}{2}e^{-S}.
\end{aligned}
\end{equation}
In general, the probability of exactly $n_i$ beam-induced excitations is
approximately

\begin{equation}
  P_i(\epsilon_b, n_i)
  \sim
  \frac{S(\epsilon_b)^{n_i}}{n_i!}
  e^{-S(\epsilon_b)}.
  \label{eq:Pi}
\end{equation}
Thus, we see that the probability $P_i$ can be written purely in terms of $S$.

\begin{figure}
  \centering
  \includegraphics[width=.9\textwidth]{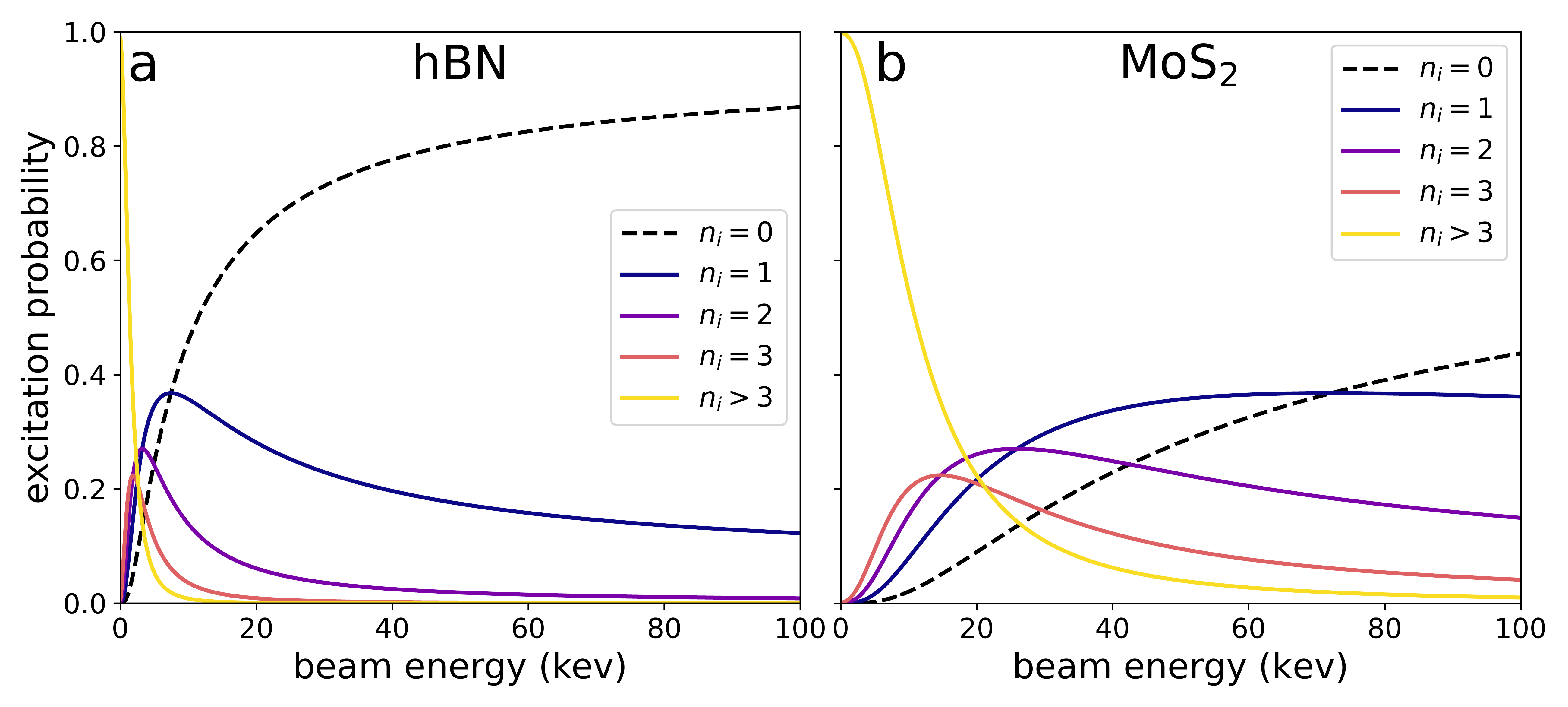}
  \caption{
    Probabilities of initially exciting a certain number of electrons $n_i$ in
    (a) hBN and (b) MoS$_2$ with a beam electron.
    The excitation probabilities tend to decreases with increasing beam energy.
    The excitation probabilities in MoS$_2$ are larger than those in hBN
    because MoS$_2$ has a significantly smaller band gap.
  } 
  \label{fig:Pi}
\end{figure}

We can now use formula (\ref{eq:Pi}) to calculate excitation probabilities in
hBN and MoS$_2$ (figure \ref{fig:Pi}).
DFT is used to obtain the plane-wave coefficients $C_{\mathbf{G}_2
+\mathbf{k}}^n$ and $C_{\mathbf{G}_3+\mathbf{k'}}^{n'}$ and eigenvalues
$\epsilon_{n\mathbf{k}}$ and $\epsilon_{n'\mathbf{k'}}$ for the pristine unit
cell of each material.
These are plugged into equation (\ref{eq:ampCode}) to obtain the amplitude for
each transition.
We sum over the squares of all resulting amplitudes to obtain $S$, which is
then plugged into formula (\ref{eq:Pi}) to obtain $P_i$ for both materials.
We emphasize that the only DFT calculation needed to determine $P_i$ is the
electronic structure relaxation of a pristine unit cell, a very inexpensive
calculation.

The probabilities plotted in figure \ref{fig:Pi} reveal some notable
trends.
First, for sufficiently large beam energies, $P_i(\epsilon_b, n_i)$ decreases
with increasing $\epsilon_b$ for all $n_i>0$.
This is because a faster beam electron has less time to interact with the
material and cause an excitation.
In this regime, $P_i(\epsilon_b)$ is proportional to $\epsilon_b^{-1}$, a
relationship originally predicted by Bethe \cite{Bethe1930, Inokuti1967,
Kretschmer2020}.
This means that multiple excitations are more likely at low beam energies.
Furthermore, the probability of remaining in the ground state $P(\epsilon_b,
n_i=0)$ vanishes as $\epsilon_b$ goes to zero.
This implies that a stationary electron in the vicinity of a material is
guaranteed to interact with the material's electrons and affect its electronic
structure.
However, the validity of formula (\ref{eq:Pi}) diminishes as $\epsilon_b$
approaches zero.
In the case of a slow beam electron, the interaction between the beam and
material electrons can no longer be approximated by the single virtual photon
transfer processes depicted in figure \ref{fig:tu}, as the amplitudes for
higher-order processes become more significant.
The effects of processes beyond the tree-level should be the subject of future
work.
Lastly, the excitation probability is inversely proportional to the material's
band gap.
This is because the zeroth component of the momentum transfer $q$ depicted in
figure \ref{fig:tu} is the difference in energy eigenvalues between the
occupied and unoccupied states (section \ref{sec:crystal}).
Thus, the smallest possible denominator of the $t$-channel in equation
(\ref{eq:M}) is proportional to the difference in eigenvalues squared.
The experimentally measured band gap of MoS$_2$ about 1.9 eV \cite{Gusakova2017}
while that of hBN is about 6.1 eV \cite{Elias2019}.
This means MoS$_2$ hosts transitions with smaller eigenvalue differences,
making the summands in equation (\ref{eq:ampCode}) larger.

With $P_i(\epsilon_b, n_i)$ derived in formula (\ref{eq:Pi}) and
$d\sigma/dE(\epsilon_b, E)$ defined in equation (\ref{eq:MF}), we now have two
of the three functions depicted in figure \ref{fig:triad} needed to calculate
the sputtering cross section in equation (\ref{eq:totSigma}).
The final ingredients are $P_f(E, \tau, n_i, n_f)$, the probability that $n_f$
excitations survive the displacement event given $n_i$ initial beam-induced
excitations, and $\{E_d\}$, the set of all displacement thresholds for all
$n_f$.
The derivations of these objects are described in the next section.

\section{Sputtering cross section}
\label{sec:sputtering}

We have demonstrated how the interaction between the beam and material
electrons can induce electronic excitations in the material.
In this section, we show how these excitations bring about much larger
sputtering cross sections than those predicted by a ground state theory.
We start by showing how electronic excitations can reduce the displacement
threshold $E_d$.
We then show that longer excitation lifetimes increase $P_f$ for nonzero $n_f$,
giving the beam-induced excitations more opportunities to lessen $E_d$.
This motivates us to write the sputtering cross section in equation
(\ref{eq:totSigma}) analytically in terms of the excitation lifetime $\tau$.
Finally, the resulting equation is used to predict the sputtering rates of
boron and nitrogen in hBN and sulfur in MoS$_2$, which can be made to agree
well with experiment for appropriate values of $\tau$ for each material.

\subsection{Effect of excitations on the displacement threshold}
\label{sec:assumptions}

We begin by describing how beam-induced excitations can reduce the
displacement threshold $E_d$ in a process called \textit{bond softening}.
As $E_d$ is the lower bound of integration over $E$ in equation
(\ref{eq:totSigma}), and the differential cross section in equation
(\ref{eq:MF}) behaves like $E^{-2}$ for small $E$, reductions in $E_d$ can
greatly increase the sputtering cross section.
Exactly how excitations change $E_d$ is an ambitious study on its own,
requiring a careful consideration of the excited electrons' evolution and
various relaxation pathways \cite{Lingerfelt2019a}.
Here instead, we make three simplifying assumptions that allow us to calculate
$E_d$ with only ground state DFT.

Before describing these assumptions, we first define some important terms.
Consider the moment immediately after the beam electron collides with a
material nucleus.
The nucleus now has a velocity corresponding to the kinetic energy
$E$ transferred from the beam electron.
The resulting nuclear motion away from its equilibrium position causes the
energy of the system to increase.
In this sense, the system climbs an energy \textit{surface} from the bottom of its
equilibrium well.
Far away from the well's bottom, the energy surface eventually plateaus.
If the system reaches this plateau, the displaced PKA moves
freely away from its initial site without deceleration. At this point, we
consider the PKA to have sputtered. We call the energy at the well's bottom
$E_0$ and that at the plateau $E_s$.
If the energy surface is static throughout the entire process, then the
displacement threshold is simply $E_d$ = $E_s$ - $E_0$.
When $E>E_d$, the system has enough energy to climb out of the well,
and the PKA sputters.  Our task now is to determine how beam-induced
excitations in the material electrons and their subsequent relaxation affect
$E_0$ and $E_s$.  To facilitate this, we make the following assumptions.

\medskip\noindent
\textit{Assumption 1: the excited electrons and holes occupy the band edges}.
To justify this, we shift our focus to the material electrons immediately after
the collision.  There are now $n_i$ electrons in the conduction band and $n_i$
holes in the valence bands.  Kretschmer et al. simulated the time-evolution of
these excitations using Ehrenfest dynamics \cite{Kretschmer2020,Li2005}.
They found that the excited electrons relax nonradiatively to the conduction
band minimum (CBM) in a few femtoseconds, while the holes in the valence band
take a similar amount of time to relax to the valence band maximum (VBM).
In contrast, the PKA takes several hundreds of femtoseconds to fully sputter
\cite{Yoshimura2018}.
Thus, the nonradiative relaxation of the electronic structure is essentially
instantaneous, and we can assume that all excited electrons and holes occupy
the CBM and VBM respectively before the PKA has been displaced by an
appreciable amount.

These findings greatly simplify the calculation of $E_0$.
Given $n_i$ excitations, $E_0$ is just the energy of the pristine system with
$n_i$ electrons and holes in the CBM and VBM respectively.
In the large crystal limit, this amounts to adding the pristine band gap $E_g$
to the system's ground state energy $n_i$ times.
In other words, $E_0(n_i) = E_0(0) + n_iE_g$.
This approximation of course ignores any binding energy between the electron
and hole.
However, we believe that this formalism allows for an efficient treatment of
the lowest order effects of excitation on $E_0$.

\medskip\noindent
\textit{Assumption 2: the sputtered system is in its ground state}.
Finding the plateau energy $E_s$ can be a bit more involved. To calculate it
properly, one must track how the excitations in the CBM and VBM evolve as the
PKA moves away from the crystal. This would require Ehrenfest dynamics of a
supercell over timescales of hundreds of femtoseconds, which is prohibitively
expensive.
Here we seek a much less costly set of DFT calculations that can still provide
a reasonable approximation for $E_s$.
To this end, we draw upon another finding of Kretschmer et al.
\cite{Kretschmer2020}.
After excitation, the displacement of the PKA causes the occupied and
unoccupied CBM and VBM states to converge into the band gap and localize on the
resulting defect.
These converging states are the bonding and antibonding states that connect the
PKA to the host crystal.
Thus, we should consider how the beam-induced excitations affect the electronic
structure on both the PKA and the remaining vacancy.

\begin{figure}
  \centering
  \includegraphics[width=\textwidth]{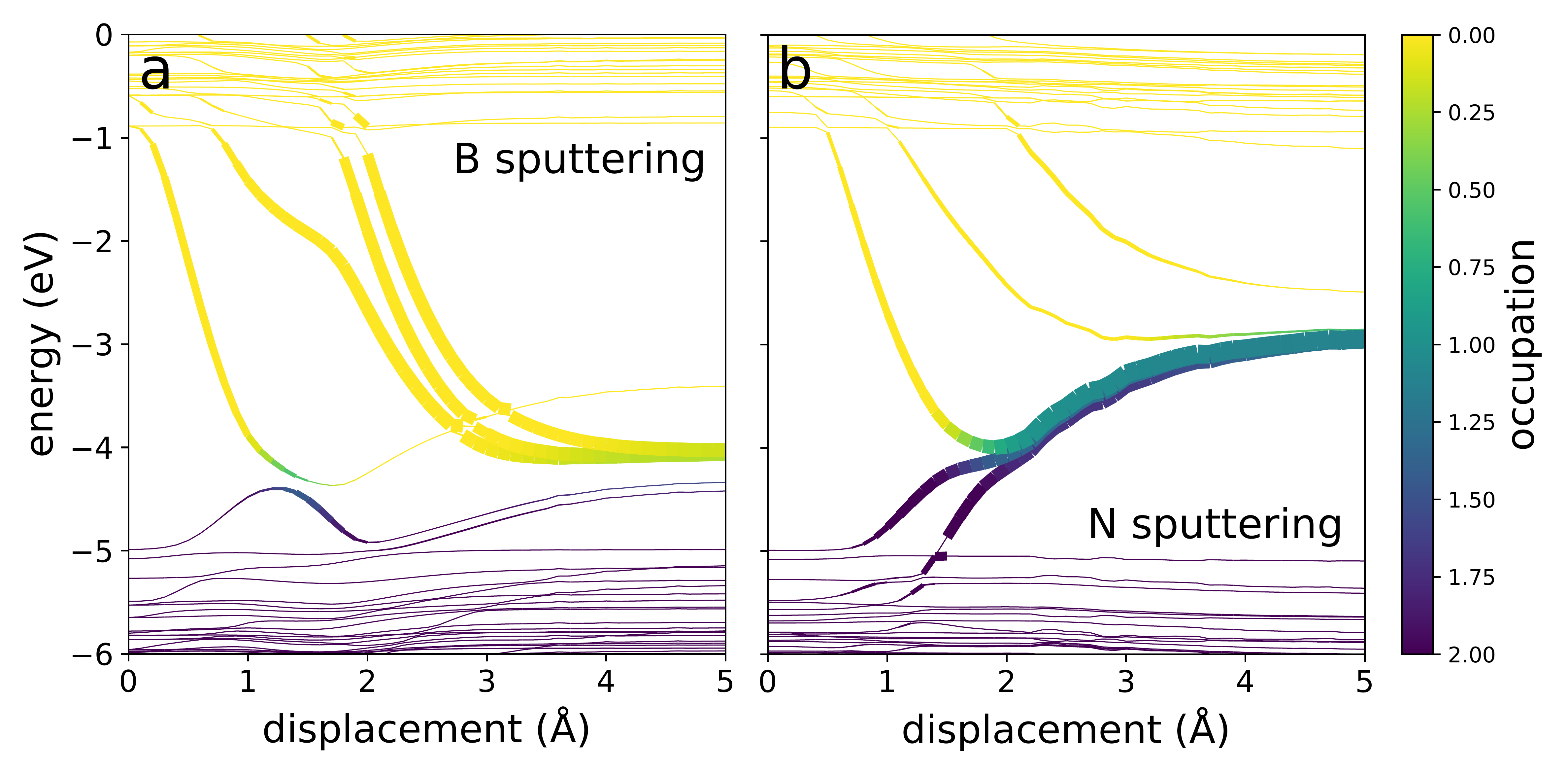}
  \caption{
    Eigenvalues and occupations of the eigenstates of armchair hBN throughout
    the sputtering process with respect to the distance between sputtered (a)
    boron and (b) nitrogen and their pristine sites.
    The thickness of each curve is proportional to the localization of the
    corresponding state on the sputtered atom.
    In hBN, the electronegative nitrogen atoms borrow charge from the
    neighboring boron atoms.
    Thus, the p-orbital states that would be occupied on an isolated boron atom
    reside in the conduction band of hBN and descend into the gap as boron
    sputters.
    On the other hand, most of the p-orbital states of nitrogen lie in the
    valence band and rise into the gap during nitrogen sputtering.
  }
  \label{fig:eigenvals}
\end{figure}

We start by considering how the ground state eigenvalues evolve as the PKA
moves away from the crystal.  For hBN, the evolution of eigenvalues differ
depending on whether boron or nitrogen is sputtered (figure
\ref{fig:eigenvals}).  Boron is electropositive while nitrogen is
electronegative.  Thus, in an hBN crystal, the nitrogen atoms borrow negative
charge from the neighboring boron atoms.  This means that the p-orbital states
that would be occupied on an isolated boron atom are vacant, hovering in
the conduction band of hBN.  Conversely, the p-orbital states of nitrogen lie
occupied in the valence band.
For both atoms, separation from the crystal causes those states to converge to
degenerate p-orbitals on their respective atoms.
When boron is sputtered, these states come from the conduction band.
This means that excited electrons residing in the CBM tend to transfer
negative charge to the sputtered boron.  However, this charge transfer is quite
energetically unfavorable since boron is electropositive. The energy cost of
this is much larger than the band gap of hBN, compelling any excess electrons
on boron to relax to the host crystal. On the other hand, the states that
localize on sputtered nitrogen must rise from the valence band.
Thus, the holes at the VBM tend to transfer positive charge to the sputtered
nitrogen, which is again energetically unfavorable for the electronegative
atom.
Thus, because charge transfer to the sputtered PKA has such a large energy
cost, we presume that in most cases, the PKA becomes charge neutral by the time
sputtering has occurred.

Furthermore, because the states that localize on the PKA converge to the same
degenerate p-orbital, the PKA assumes its ground state after it has sputtered,
regardless of how many electrons the beam excites.
We call the energy of the isolated ground state atom $E_a$.
Meanwhile, the convergence of the bonding and antibonding states localized on
the remaining vacancy during the displacement causes the excited electron and
hole energy levels to cross multiples times.  This means that there are plenty
of chances for an excited electron to relax nonradiatively as the PKA separates
from the material.  Thus, we assume that the host crystal also relaxes to its
ground state for most sputtering events.

Lastly, the sputtering threshold is always larger than the vacancy formation
energy, because some of the energy transferred to the PKA disperses to the
neighboring atoms.
For this reason, we calculate the energy of the vacant system without relaxing
the structure, as it has been shown that the free energy gained by leaving the
structure unrelaxed roughly matches the energy dissipated to the
surrounding material \cite{Komsa2012}.
We call the energy of the unrelaxed ground state vacancy $E_v$.
Thus, we calculate the plateau energy using $E_s = E_v + E_a$, the sum of the
ground state vacancy and isolated atom free energies.
\medskip\noindent
\textit{Assumption 3: the PKA travels at a constant velocity}.
Each number of excitations creates its own energy surface.  In this way,
electronic excitation and relaxation cause the system to \textit{hop} from
surface to surface.  In our problem, the initial beam-induced excitation
perches the system on an elevated surface.  Once there, electronic relaxation
via spontaneous emission (SE) enables downward surface hopping at the cost
of the emitted photon energy.
In this framework, $E_d$ is the energy gained along all energy surfaces
that the system traverses during sputtering.  For example, suppose the system
emits one photon as it sputters.  $E_d$ is then the energy gained along the
portion of the excited energy surface traversed before emission plus that
gained along the relaxed surface after emission.
This is how SE can change $E_d$.
However, the effect of SE on $E_d$ must diminish as the PKA moves further away
from the crystal.
This means there must be some distance $d$ beyond which SE no
longer affects the forces on the sputtered atom, leaving the energy surface
unchanged.
We take $d=4.5$ {\AA} in line with Kretschmer et al.'s study, though we
acknowledge that the exact meaning of this distance is different in their work
\cite{Kretschmer2020}.

We wish to capture this idea while making the calculation of $P_f$ as
intuitive as possible.
We do this by assuming that each energy surface is constant except for a step
when the PKA reaches a distance $d$ away from its equilibrium site.
This means that immediately after the collision that excites $n_i$ electrons,
the surface assumes a constant energy $E_0(n_i)$.
Each relaxation via SE causes the system to drop to the surface beneath it,
decreasing its energy by $E_g$.
We define $n_f$ as the number of electronic excitations that survive the PKA's
traversal of distance $d$ to the energy step.
Thus, the energy of the system just before the PKA reaches the step is
$E_0(n_f)$,
Beyond the step, all energy surfaces have energy $E_s$, and SE no longer
changes the energy surface. 
With this formalism, $E_d$ is simply the height of the step,

\begin{equation}
  E_d(n_f) = E_s - E_0(n_f) = E_v + E_a - E_0(0) - n_f E_g.
  \label{eq:Ed}
\end{equation}

Thus we have derived a simple relationship for how electronic excitations
affect $E_d$.
However, before we show how this relationship affects the sputtering
cross section, we must first point out an important caveat.
Equation (\ref{eq:Ed}) suggests that $E_d$ can be negative for large $n_f$.
In these cases, the affected atom accelerates away from its pristine site even
without energy transfer from the beam electron.
This would seem to suggest that the sputtering cross section is infinite, since
the PKA would sputter for any energy transfer, no matter how small.
This is corroborated by the fact that the integral in equation
(\ref{eq:totSigma}) diverges as $E_d$ approaches zero, meaning that the
sputtering cross section $\sigma$ would approach infinity.
However, this is nonsensical, as it would imply an infinite sputtering rate for
a finite beam current.
In reality, even though the beam electron does nudge all of the material nuclei
to some extent, one nucleus always receives a bigger nudge than the rest.
This is the atom onto which the beam-induced excitations localize.
Thus, for each beam electron, only one atom's displacement threshold is reduced
by electronic excitations.
For this reason, the displacement cross section should never exceed the
cross sectional area occupied by a single target atom.
This means that $E_d$, the lower integration bound in equation
(\ref{eq:totSigma}), must have a lower bound itself, which we call
$E_\text{min}$.
When $E_d$ falls below $E_\text{min}$, we replace it with $E_\text{min}$.
We can approximate $E_\text{min}$ by setting the displacement cross section in
equation (\ref{eq:basicSigma}) equal to the area occupied by the PKA and
solving for the displacement threshold (section \ref{app:Emin}).
With this, we are finally ready to use equation (\ref{eq:Ed}) to determine the
set of displacement thresholds $\{E_d\}$ to insert into equation
(\ref{eq:totSigma}).

All positive displacement thresholds computed for this work are listed in table
\ref{tab:Ed}.
Explanations for the calculations of $E_v$, $E_a$, $E_0(0)$, and $E_g$ are
given in subsctions \ref{sec:hBN} and \ref{sec:MoS2}.
The three largest displacement thresholds for MoS$_2$ with $n_f = 0$, 1, and 2
excitations are similar to those calculated with DFT-based molecular dynamics
simulations \cite{Kretschmer2020}.
This provides some assurance that our simplified approach to calculating $E_d$
yields reasonable results.
Excitation numbers $n_f$ greater than those listed in table \ref{tab:Ed} make
$E_d$ negative, in which case the exact value of $E_d$ is not important since
it is replaced with $E_\text{min}$ in the calculation of $\sigma$.
With that said, it is critical to consider large $n_f$, even if the resulting
$E_d$ is less than $E_\text{min}$.
These large $n_f$ can make appreciable contributions to the total
cross section, especially at small beam energies for which $P_i(\epsilon_b,
n_i)$ is significant for large $n_i$ (figure \ref{fig:Pi}).
As $n_f$ cannot exceed $n_i$, this means that we must consider sufficiently
large $n_i$ to acknowledge these contributions.
Eventually, these contributions diminish as we increment $n_i$, because the
$n_i!$ in the denominator of $P_i$ in formula (\ref{eq:Pi}) eventually
outgrows the $S^{n_i}$ in the numerator.
We can therefore truncate the summation over $n_i$ in equation
(\ref{eq:totSigma}) at some adequately large $n_i^\text{max}$.
This means that $n_i^\text{max}$ must be converged for each material, as
materials with greater $S$ require greater $n_i^\text{max}$.
In this work, we choose $n_i^\text{max}$ large enough so that including more
excitations increases the sputtering cross section by less than 1\% for the
smallest experimental beam energy considered for each matieral (figure
\ref{fig:nimax}).

Thus, we have shown how $E_d$ depends only on the number of surviving
excitations $n_f$.  We must now determine the likelihood that $n_f$ excitations
survive given $n_i$ beam-induced excitations.
This is done by comparing the excitation lifetime $\tau$ to $t_s$, the time it
takes for the PKA to travel a distance $d$.
We handle this task in the next subsection, where we write the sputtering
cross section in equation (\ref{eq:totSigma}) in terms of
$\tau$.

\begin{table}
  \centering 
  \begin{tabular}{ccccc}
    \toprule &
    \multicolumn{4}{c}{displacement threshold (eV)} \\
    \cmidrule{2-5}
    $n_f$ &0 &1 &2 &3 \\
    \midrule
    B from hBN     &12.85 &8.78 &4.71 &0.64 \\  
    N from hBN     &12.71 &8.64 &4.57 &0.50 \\  
    S from MoS$_2$ &6.92  &5.04 &3.16 &1.28 \\  
    \bottomrule
  \end{tabular}
  \caption{
    All computed positive displacement thresholds for sputtering from the hBN
    armchair edge and MoS$_2$ surface.
    The thresholds decrease as the number of surviving excitations $n_f$
    increases.
  } 
\label{tab:Ed}
\end{table}

\subsection{Sputtering cross section in terms of the excitation lifetime}
\label{sec:lifetime}

In hBN and MoS$_2$, occupation of the antibonding state localized on the PKA
does not affect the free energy of the sputtered system, since the antibonding
and bonding states to converge to the same degenerate p-orbital in the limit
that the PKA is isolated from the host material.
Therefore, only the excitation lifetime of the host material needs be
considered in calculating $P_f(n_i, n_f)$, the probability that $n_f$ of the
$n_i$ beam-induced excitations survive long enough to reduce the displacement
threshold.

We define the ratio of surviving excitations as

\begin{equation}
  R(E, \tau) = e^{-t_s(E)/\tau},
  \label{eq:R}
\end{equation}
where $t_s$ is the time it takes for the sputtered atom to travel a distance
$d$, and the excitation lifetime $\tau$ is determined for each material by
fitting the cross section to experimental data.
Given $n_i$ beam-induced excitations, the probability that $n_f$ excitations
survive is

\begin{equation}
  \begin{aligned}
    P_f(n_i, n_f)
    &=
    \begin{pmatrix}
      n_i\\n_f
    \end{pmatrix}
    R^{n_f}(1-R)^{n_i - n_f}
    =
    \begin{pmatrix}
      n_i\\n_f
    \end{pmatrix}
    \sum^{n_i-n_f}_{n=0}
    \begin{pmatrix}
      n_i-n_f\\n
    \end{pmatrix}
    (-1)^n R^{n_f+n}.
  \end{aligned}
  \label{eq:Pf}
\end{equation}
%
Equation (\ref{eq:Pf}) allows us to rewrite the integral in equation
(\ref{eq:totSigma}) as

\begin{equation}
  \begin{aligned}
    &\int_{E_d}^{E_\text{max}}
    P_f \frac{d\sigma}{dE} dE
    =
    \begin{pmatrix}
      n_i\\n_f
    \end{pmatrix}
    \sum^{n_i-n_f}_{n=0}
    \begin{pmatrix}
      n_i-n_f\\n
    \end{pmatrix}
    (-1)^n
    \int^{E_\text{max}}_{E_d}
    R^{n_f+n}
    \frac{d\sigma}{dE}
    dE.
  \end{aligned}
  \label{eq:integral}
\end{equation}
Assumption 3 from the previous subsection again aids us here.
Because the PKA travels at a constant velocity until it reaches the sputtering
distance $d$, we can write $t_s$ from equation (\ref{eq:R}) quite simply as
$t_s = d\sqrt{M/2E}$.
This allows for the straightforward analytical integration of equation
(\ref{eq:integral}).  Using $d\sigma/dE$ defined in equation (\ref{eq:MF}), the
integral on the right can be written as

\begin{equation}
  \begin{aligned}
    \int R^{n_f+n} \frac{d\sigma}{dE} dE
    &=
    \pi\left(\frac{Z\alpha}{|\mathbf{p}|\beta}\right)^2
    \bigg[
      \frac{4\tau^2 E_\text{max}}{d^2M} \frac{\mu\xi+1}{\mu^2} e^{-\mu\xi}
      \\&+
      2\beta(\pi Z\alpha+\beta)\text{Ei}(-\mu\xi)
      +
      \frac{\pi Z\alpha\beta\tau}{\mu d}\sqrt{\frac{8E_\text{max}}{M}}
      e^{-\mu\xi}
    \bigg]
    +\text{const}.,
  \end{aligned}
  \label{eq:analytical}
\end{equation}
where we define

\begin{equation}
  \xi \equiv \frac{-d}{\tau} \sqrt{\frac{M}{2E}}
  \quad\text{and}\quad
  \mu \equiv n_f + n,
  \label{eq:xi}
\end{equation}
and the function Ei is the exponential integral,

\begin{equation}
  \text{Ei}(x) = \int_{-\infty}^x \frac{e^t}{t}dt.
  \label{eq:Ei}
\end{equation}
Looking back at equation (\ref{eq:totSigma}), we now have everything we need to
evaluate the sputtering cross section.
We derived $P_i(\epsilon_b, n_i)$ in formula (\ref{eq:Pi}), and we performed
the integration over $E$ analytically in equations (\ref{eq:integral}) and
(\ref{eq:analytical}), setting $\{E_d\}$ in equation (\ref{eq:Ed}) as the lower
integration bounds.
We also have a criterion to truncate the summation over $n_i$ at
$n_i^\text{max}$ for a given material, as described at the end of section
\ref{sec:assumptions}.
In the following subsections, we use these results to calculate the sputtering
cross sections of hBN and MoS$_2$.

\subsection{Boron and nitrogen sputtering from hexagonal boron nitride}
\label{sec:hBN}

Electron beam irradiation has been shown to bore nanopores in monolayer hBN at
beam energies far beneath the calculated ground state critical energy of
$\epsilon_c\sim80$ keV \cite{Jin2009,Meyer2009,Kotakoski2010,Cretu2015}.
In a pristine hBN layer, these beam-induced pores can be initialized from
isolated boron and nitrogen vacancies.
The atoms surrounding these vacancies have reduced coordination numbers, and
thus, smaller displacement thresholds than those in the pristine material.
Therefore, the atoms lining the defect are more likely to sputter than their
surface counterparts for a given beam energy.
As the edge atoms continue to sputter away at a high rate, the nanopore grows,
eventually extending up to a few nm in diameter
\cite{Meyer2009,Cretu2015,Dogan2020}.

\begin{figure} 
  \centering
  \includegraphics[width=\textwidth]{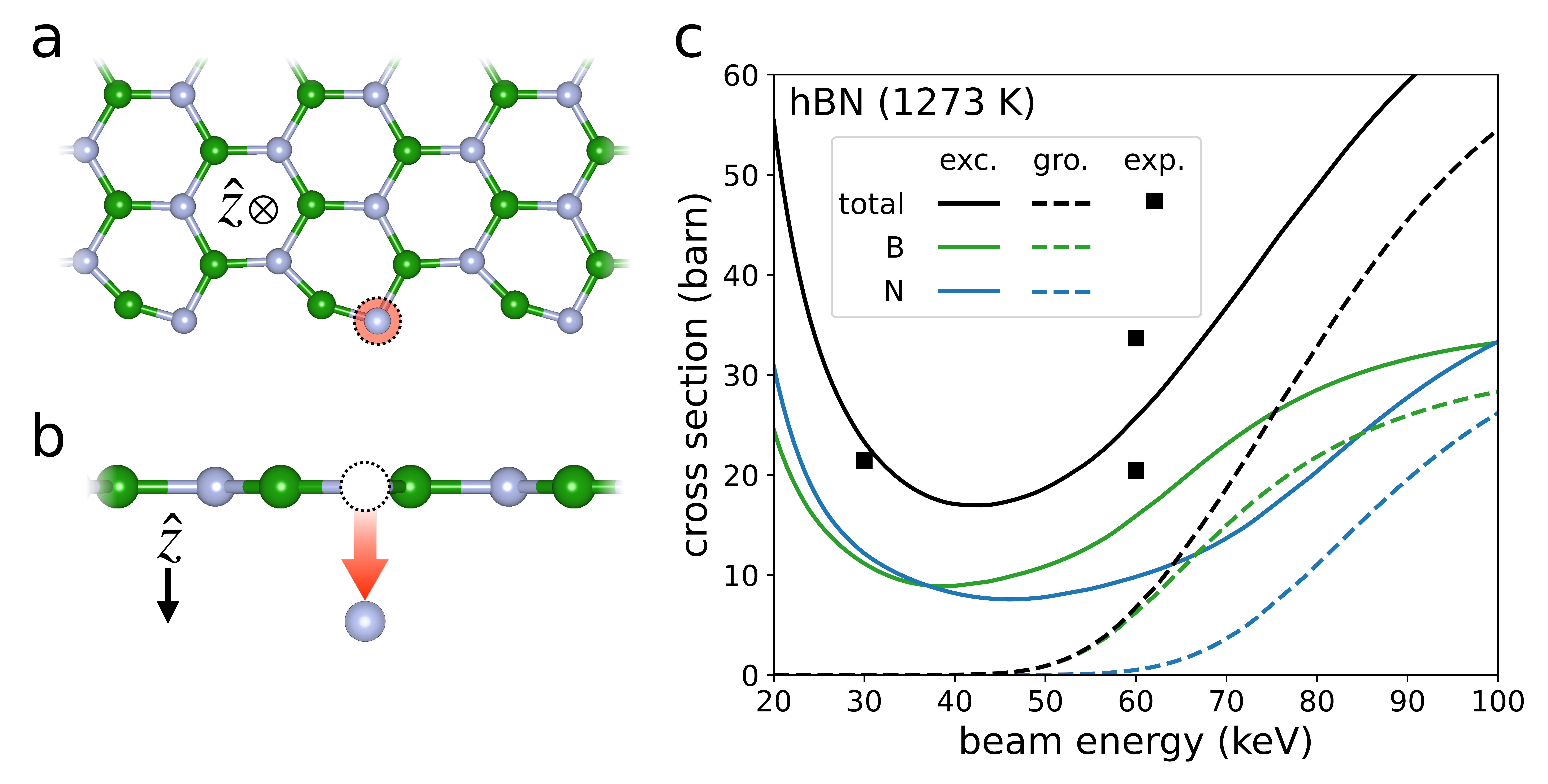}
  \caption{
    Sputtering from the armchair edge of hBN.
    The schematics in panels (a) and (b) depict the sputtering of nitrogen from
    out-of-plane and in-plane perspectives respectively.
    Panel (c) plots the sputtering cross sections of both boron and nitrogen in
    green and blue respectively.
    The black curves are the total sputtering cross section, i.e., the sum of
    their corresponding blue and green curves.
    The solid curves account for beam-induced excitations assuming an
    excitation lifetime of $\tau=240$ fs (labelled ``exc.'' for excited).
    The dashed curves ignore the possibility of beam-induced excitation
    (labelled ``gro.'' for ground).
    The black squares are experimentally observed sputtering cross sections
    for hBN at 1273 K \cite{Cretu2015}.
    The consideration of beam-induced excitations reduces the disagreement
    between theory and experiment substantially.
  }
  \label{fig:edgeCross}
\end{figure}

Cretu et al. measured the radial growth of these nanopores under electron
irradiation for several temperatures ranging from 673 to 1473 K
\cite{Cretu2015}.
By dividing the radial growth rate by the beam current, they estimated the
sputtering cross section to be around 25 barn under both 30 and 60 keV beams at
temperatures of 1273 K and below (figure \ref{fig:edgeCross}c).
The cross sections were fairly temperature independent at these relatively low
temperatures.
Cretu et al. also found that the edges of these pores most often assume an
armchair structure.
Therefore, in an attempt to reproduce these measurements, we calculate the
cross section of boron and nitrogen sputtering from an armchair edge at 1273 K
(figure \ref{fig:edgeCross}).
To evaluate the displacement thresholds defined in equation (\ref{eq:Ed}) and
listed in table \ref{tab:Ed}, $E_v$ is the free energy of the armchair edge
supercell with a single boron or nitrogen vacancy, while $E_a$ is that of an
isolated boron or nitrogen atom.
$E_0(0)$ and $E_g$ are then the calculated free energy and band gap of the
pristine armchair edge supercell.
We plug the resulting set of displacement thresholds $\{E_d\}$ into equation
(\ref{eq:totSigma}) to calculate the sum of the boron and nitrogen sputtering
cross sections.
The excitation probabilities of hBN plotted in figure \ref{fig:Pi}a are
then used for $P_i$.
Strictly speaking, the calculation of $P_i$ should consider the effects of the
edge state orbitals.
However, in our formalism, the beam electron is in a momentum eigenstate that
is highly delocalized in real space.
We therefore assume that the radius of the beam is much larger than that of the
nanopore.
This means that the majority of the beam-matter interactions occur in regions
of pristine material, validating the use of the $P_i$ calculated for pristine
hBN.
Future work should consider the effects of localized beam electron states to
simulate beams with smaller focal points.
Lastly, temperature effects on the cross section are considered in the manner
described in our previous work \cite{Yoshimura2018}.

Fitting our cross section curves to the data of Cretu et al. yields an excellent
agreement if the predicted excitation lifetime is set to $\tau \sim 240$ fs.
This predicted lifetime is much shorter than the reported excitation lifetime
of $\sim$0.75 ns in pristine hBN \cite{Li2016b}, indicating that the sputtering
process can significantly reduce the excitation lifetimes of hBN.
We suspect that the atomic motion gives rise to non-radiative relaxation
pathways that are not explicitly accounted for here. 
This motivates a closer investigation into the full electronic evolution of
the hBN system post-collision.
However, such a study is beyond the scope of this work.
With that said, the electronic structure in the vicinity of a sputtering PKA
differs significantly from that of the pristine room-temperature systems in
which excitation lifetimes are experimentally measured.
There is therefore no reason to expect the predicted lifetimes in this
work to match those obtained by experiment.

We also see that the sputtering cross section, and thus the growth rate of the
nanopore, is minimized for beam energies between 30 and 60 keV.
Below these energies, the sputtering cross section begins to grow as the beam
energy decreases.
This is due to a non-negligible probability of final excitation numbers $n_f >
3$, for which $E_d(n_f)$ falls below $E_\text{min}$ (table \ref{tab:Ed}).
In these cases, the sputtering cross section is $\Omega \sim 5\times10^8$ barn,
seven orders of magnitude larger than the measured cross section at 30 keV.
This suggests that one can expect the beam-induced nanopores in hBN to grow under
beam energies as low as 1 keV.
However, one must again be cautious of the predicted cross section at low beam
energies in which the tree-level theory starts to break down.
Nevertheless, figure \ref{fig:edgeCross}c demonstrates a strong beam-energy
dependence in the sputtering cross section across a wide range of
experimentally relevant beam-energies.
These findings could facilitate precise control of nanopore growth rates in hBN
under electron irradiation.

\subsection{Sulfur sputtering from molybdenum disulfide}
\label{sec:MoS2}

\begin{figure}
  \centering
  \includegraphics[width=.95\textwidth]{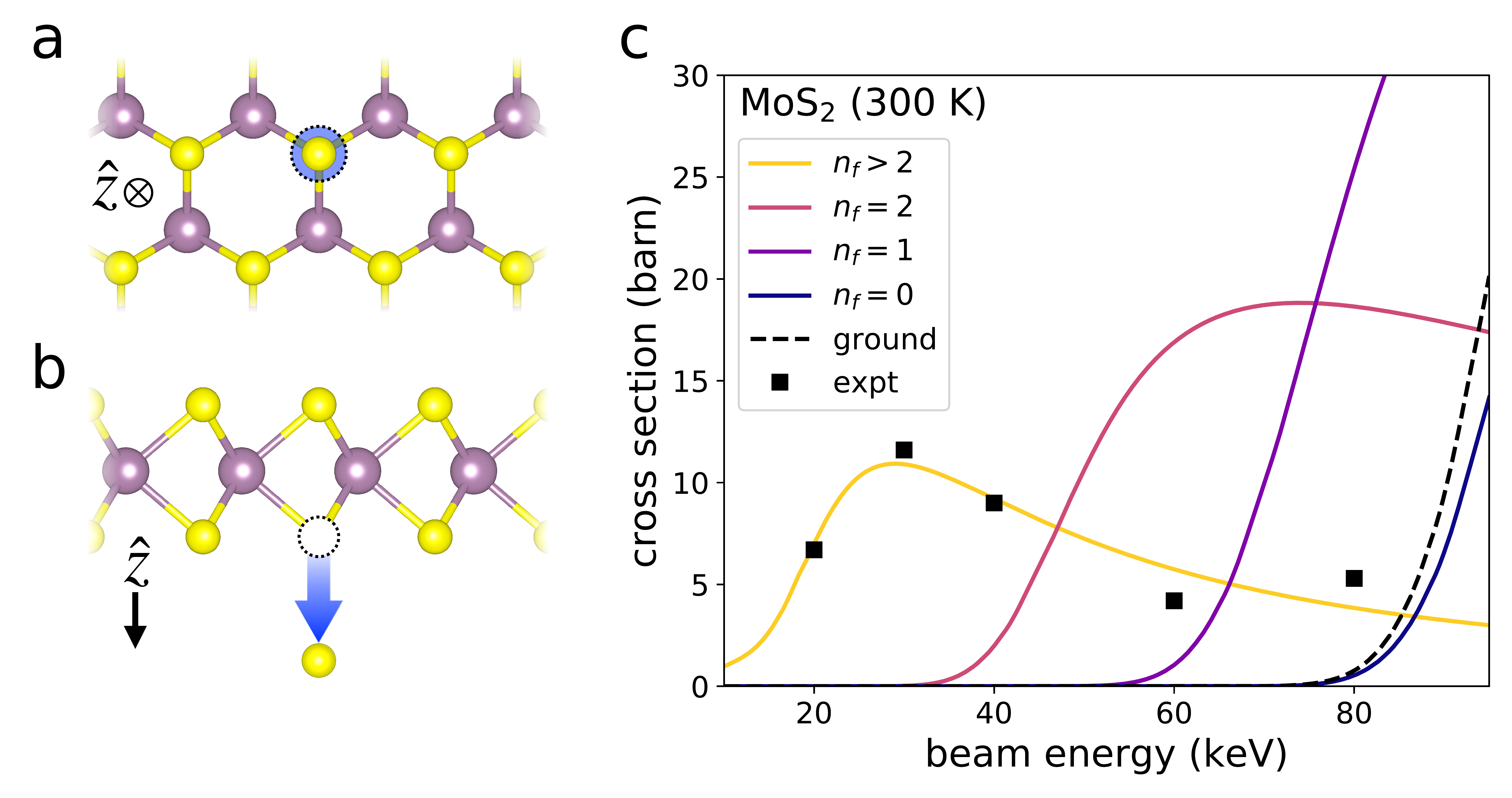}
  \caption{
    Sputtering of sulfur from the outgoing MoS$_2$ surface.
    The sputtering of molybdenum from MoS$_2$ can be ignored because its
    displacement threshold is significantly larger than that of sulfur.
    The schematics in panels (a) and (b) depict the sputtering process from
    out-of-plane and in-plane perspectives respectively.
    Panel (c) plots the contributions of various numbers of final excitations
    $n_f$ to the sputtering cross section assuming an excitation lifetime of
    $\tau=81$ fs.
    The black dashed curve plots the predicted cross section ignoring the
    possibility of beam-induced excitation.
    The black squares are experimentally observed cross sections at 300 K
    \cite{Kretschmer2020}.
    The contribution from the sum of all $n_f>2$ final excitations matches
    the experimental cross section remarkably well.
  }
  \label{fig:MoS2Cross}
\end{figure}

Kretschmer et al. measured the sputtering cross section of sulfur from MoS$_2$
for several beam energies ranging from 20 to 80 keV \cite{Kretschmer2020}.
They found a peak in the cross section at 30 keV, much less than the predicted
ground state $\epsilon_c\sim 90$ keV \cite{Kretschmer2020}.
To help explain this unexpected peak, we calculate the cross section for sulfur
sputtering from pristine MoS$_2$ (figure \ref{fig:MoS2Cross}).
The vacant system free energy $E_v$ for the calculation of $\{E_d\}$ is that of a
MoS$_2$ supercell with a single sulfur vacancy.
$E_a$ is then the free energy of an isolated sulfur atom, and $E_0(0)$ is that
of the pristine MoS$_2$ supercell.
We then set $E_g$ equal to the experimental band gap of 1.88 eV
\cite{Gusakova2017}.
Using the resulting $\{E_d\}$, we find that summing over the contributions
of final excitation numbers $n_f > 2$ to the sputtering cross section produces
a peak just below 30 keV, matching Kretschmer et al.'s findings remarkably
well.
In fitting to this peak, we predict an excitation lifetime of $\tau \sim 81$
fs.
This is again much shorter than the excitation lifetime of pristine MoS$_2$,
which is on the order of a few picoseconds \cite{Korn2011, Lagarde2014,
Palummo2015a}.
However, this is also much shorter than the fitted lifetime of hBN found in the
previous subsection, consistent with the fact that the excitation lifetime of
pristine hBN is much longer than that of MoS$_2$.

The difference between the two materials' lifetimes leads to markedly
dissimilar cross section behaviors at low beam energies.
Below beam energies of 30 keV, the cross section of MoS$_2$ gradually drops to
zero with decreasing $\epsilon_b$.
In contrast, hBN's total cross section has a minimum at around 40 keV and
begins to increase as $\epsilon_b$ decreases.
Eventually, hBN's sputtering cross section peaks before dropping quickly to
zero as the beam energy goes to zero (figure \ref{fig:edgePeaks}).
However, this peak occurs at around 0.5 keV, far below the lower energy bound
of figure \ref{fig:edgeCross}c.
These distinct cross section behaviors can be explained by the amplified
sensitivity of $n_f$ to $\tau$ at low beam energies.
Equations (\ref{eq:R}) and (\ref{eq:Pf}) tell us that larger $\tau$
makes large $n_f$ more likely.
At the same time, $n_f$ cannot exceed $n_i$.
Thus, $n_f$ is more sensitive to $\tau$ at low beam energies for which $n_i$ is
large.
Accordingly, because $\tau$ is greater in hBN than in MoS$_2$, the expected
values of $n_f$ in hBN are much larger than those of MoS$_2$ at low beam
energies.
This explains why the effect of considering excitations is much more
pronounced in hBN than in MoS$_2$.
Furthermore, the difference in the cross section behavior is exacerbated by the
fact that sulfur is heavier than both boron and nitrogen, so that its
post-collision velocity is always smaller for a given energy transfer $E$.
It follows that $t_s$ is larger, and thus, $P_f$ is smaller for a given $E$ in
MoS$_2$.
This again increases the likelihood that hBN has more final excitations than
MoS$_2$, meaning that the effects of beam-induced excitation on hBN's
sputtering cross section are greater.

We also plot the contributions of $n_f = 0$, 1, and 2 excitations to MoS$_2$'s
sputtering cross section separately.
In doing so, we see that the contributions of $n_f = 1$ or 2 would conceal the
peak at 30 keV. 
Thus, it seems that the likelihoods of $n_f = 1$ or 2 are somehow suppressed.
This suggests that the individual beam-induced excitations and subsequent
relaxation are in some cases correlated.  That is, the excitation and/or
relaxation rates of a given electronic transition are affected by the
coinciding distribution of electronic excitations.
Thus, a proper treatment of these excitations and their effect on the sputtering
cross section requires that the excitation probabilities of every possible
transition are calculated for every possible excitation configuration.
Such a nonlinear calculation is beyond the scope of this work, but
should certainly be pursued in a future study.
Nonetheless, the methods laid out here demonstrate that the consideration of
beam-induced excitation can provide a quantitative justification for the sulfur
sputtering rate to peak at a beam energy well-below the expected ground state
critical energy.

\section{Conclusion}
\label{sec:conclusion}

In this paper, we developed a first-principles method to more accurately
describe electron beam-induced sputtering cross sections in 2D insulating
crystals by accounting for beam-induced electronic excitations and their
subsequent relaxations.
The method combines QED scattering theory with DFT electronic structure
calculations to determine the likelihood of beam-induced excitation.
The results show that the excitation probability is inversely proportional
to both the material's band gap and beam electron's kinetic energy.
We then show how these nonzero excitation probabilities increase the predicted
sputtering cross sections of both hBN in MoS$_2$.  These cross sections can be
made quantitatively similar to those obtained experimentally by treating the
excitation lifetime as a fitting parameter.
The methods laid out are computationally efficient, requiring only a few
ground state electronic optimization calculations for each cross section curve.
Thus, the formalism that we have developed can be easily applied to any 2D
crystalline material to simulate the rates of atomic displacement under
electron irradiation.

With that said, 
several questions naturally arise from our study.
For example, why is the excitation lifetime reduced so drastically during
sputtering?
How might excitation and relaxation rates of different transitions be
correlated?
How might preexisting defects affect those rates?
These questions urge follow up work to address the full breadth of physical
processes involved in beam-matter interactions.
Future studies should also consider additional electronic relaxation pathways
to determine their effect on $P_f$.
Correspondingly, other electronic responses such as ionization, core
excitations, and second order electronic excitation effects such as Auger
scattering can be incorporated into the calculation of $P_i$
\cite{Lagarde2014,Kozawa2014,Nie2015,Shi2013b}.
Moreover, spin polarization effects can also be examined by making $\mathcal{M}$
spin-dependent, i.e., not averaging over spins as is done in equation
(\ref{eq:M}).
The beam electron path can also change significantly after collision with
the nucleus.
Ensuing research should investigate how these altered trajectories generate new
excitation probabilities $P_i$, which must certainly play a role in 3D bulk
materials.
Furthermore, the methods here can be modified to accommodate more advanced DFT
techniques.
For example, the calculation of $P_i$ should be made compatible with ultrasoft
and projector augmented wave pseudopotentials \cite{Blochl1994} in a manner
similar to that employed in modern GW codes \cite{Shishkin2006a, Gajdos2006,
Paier2005}.
In addition, the plane-wave coefficients of excited electronic states can be
calculated self-consistently using constrained DFT \cite{Kaduk2012}.  
Perhaps most importantly, progress in this field requires much more experimental
data.
We therefore hope this paper encourages new experimental investigation into
beam-induced sputtering for beam energies below the predicted ground state
critical energy.

Clearly, the combined QED-DFT approach to modeling beam-induced excitations and
their effect on the sputtering cross section opens up a rich and diverse field
of physics for both theoretical and experimental exploration.
We hope that this work and the work it may stimulate can eventually enable the
use of electron beams for precise atomic-scale engineering of crystalline
materials.

\section{Methods}
\label{sec:methods}

DFT \cite{Hohenberg1964, Kohn1965} was used to determine all electronic and
ionic structures from first-principles.
Free energy calculations were carried out with the Vienna ab initio Simulation
Package (VASP) \cite{Kresse1996, Kresse1996a} implementing the projector
augmented wave (PAW) method \cite{Blochl1994} along with the
Perdew-Burke-Ernserhof (PBE) generalized gradient approximation (GGA) to the
exchange correlation functional \cite{Perdew1996}.
Van der Waals interactions were accounted for using the optB88-vdW density
functional methods \cite{Klimes2010, Klimes2011}.
All parameters were converged so that any increase in precision would change
the total free energy by less than 1 meV per atom.
The cutoff energies for hBN and MoS$_2$ were set to 800 and 550 eV
respectively, while the BZs of both materials' pristine unit cells were sampled
with a $\Gamma$-centered $6\times6\times1$ Monkhorst-Pack meshes
\cite{Monkhorst1976}, corresponding to 0.417 and 0.328 and k-points per inverse
{\AA} for hBN and MoS$_2$ respectively.
To achieve the same k-point densities, surface vacancies in MoS$_2$ and edge
vacancies in hBN were placed in respective $6\times6\times1$ and
$4\times1\times1$ supercells whose BZs were sampled with a single k-point on
$\Gamma$.
The heights of the hBN and MoS$_2$ cells were 12 {\AA} and 20 {\AA}
respectively to provide sufficient separation from periodic images.
Nanoribbon structures were used to simulate isolated armchair edges in hBN.
These ribbons were more than 16 {\AA} across and placed in cells 28 {\AA} wide
to avoid interactions between opposing edges and periodic images.
Lastly, relaxation iterations of ionic positions and lattice constants
persisted until the all Hellmann-Feynman forces settled below 1 
meV/{\AA}.

Because equation (\ref{eq:ampCode}) relies on the orthogonality of the
Kohn-Sham orbitals, the optimized norm-conserving Vanderbilt pseudopotentials
\cite{Hamann2013, Schlipf2015} implemented in Quantum ESPRESSO
\cite{Giannozzi2009} were used to determine all plane-wave coefficients
$C^n_\mathbf{G+k}$.
The sum of all transition probabilities $S$ defined in equation (\ref{eq:S})
was used to gage the convergence of all parameters, which were deemed converged
when any increase in precision changed $S$ by less than 5\% (figure
\ref{fig:convergences}).
The parameter values that satisfy this criteria differed substantially from
those needed for free energy convergence.
The cutoff energies for hBN and MoS$_2$ were set to 299 and 286 eV
respectively, while their cell-heights were respectively set to 18 and 12
{\AA}.
For both materials, the maximum virtual photon momentum
$|\mathbf{q}_\text{max}|$ required for convergence fell well within their first
BZs.
We therefore chose $|\mathbf{q}_\text{max}|$ to be the magnitude of
high-symmetry point $K$ in each respective BZ.
The maximum number of initial excitations considered for hBN and MoS$_2$ were
$n_i^\text{max} = 5$ and 9 respectively (figure \ref{fig:nimax}).
Finally, convergence of $S$ requires extremely dense k-point sampling of the
BZ.
This necessitates fitting a curve to $S$ calculated for various k-point mesh
densities and extrapolating to an infinitely fine mesh to estimate the
converged value of $S$ (section \ref{app:fitting}).
The most dense k-point meshes used to fit these curves had dimensions of
$45\times45\times1$ and $36\times36\times1$ for hBN and MoS$_2$ respectively.

\section{Acknowledgements}
\label{sec:acknowledgements}

Calculations were performed at the Center for Computational Innovations at
Rensselaer Polytechnic Institute, Livermore Computing Center at Lawrence
Livermore National Laboratory (LLNL), and Compute and Data Environment for
Science at Oak Ridge National Laboratory (ORNL), which is supported by the
Office of Science of the U.S. Department of Energy (DOE) under Contract No.
DE-AC05-00OR22725.
This work was performed under the auspices of the U.S. Department of Energy by
LLNL under Contract DE-AC52-07NA27344.
This material is also based upon work supported by the U.S. DOE, Office of
Science, Office of Workforce Development for Teachers and Scientists, Office of
Science Graduate Student Research (SCGSR) program.
The SCGSR program is administered by the Oak Ridge Institute for Science and
Education for the DOE under contract number DE‐SC0014664.
Funding was provided by the National Science Foundation (Award 1608171).  Work
(BGS, PG, DL, JJ) was also supported by ORNL's Center for Nanophase Materials
Sciences, a U.S. DOE Office of Science User Facility.


\makeatletter\@input{maux.tex}\makeatother  

\bibliography{library}
\bibliographystyle{ieeetr}
\end{document}


\maketitle

\section{Invariant matrix element $\mathcal{M}$}
\label{app:M}

As the excitation amplitude in equation (\ref{eq:ampCode}) contains the
invariant matrix element $\mathcal{M}$ and not $|\mathcal{M}|^2$, we are unable
to use the spin sum identities typically used to derive many QED cross sections
\cite{Peskin1995, Lancaster2014}.
Nonetheless, the evaluation of $\mathcal{M}$ in equation (\ref{eq:M}) is
straightforward, though a bit cumbersome.
In the Dirac basis \cite{Bjorken1964},

\begin{equation}
  \gamma^0
  =
  \mqty( I&0 \\ 0&-I )
  \quad\text{and}\quad
  \gamma^i
  =
  \mqty( 0&\sigma^i \\ -\sigma^i&0),
  \label{eq:diracBasis}
\end{equation}
where

\begin{equation}
  I = \mqty( 1&0 \\ 0&1 )
  \quad\text{and}\quad
  \vec{\sigma}
  =
  \left\{
    \mqty( 0&1 \\ 1&0 ),
    \mqty( 0&-i \\ i&0 ),
    \mqty( 1&0 \\ 0&-1 )
  \right\}.
  \label{eq:sigmas}
\end{equation}
The electron spinors in equation (\ref{eq:M}) can be written as

\begin{equation}
  u^1(p)
    =
    \sqrt{\epsilon + m}
    \mqty(
      1 \\[2pt] 0 \\[2pt]
      \dfrac{p^z}{\epsilon + m} \\[8pt]
      \dfrac{p^x + i p^y}{\epsilon + m}
    )
  \quad\text{and}\quad
  u^2(p)
    =
    \sqrt{\epsilon + m}
    \mqty(
      0 \\[2pt] 1 \\[2pt]
      \dfrac{p^x - i p^y}{\epsilon + m} \\[8pt]
      \dfrac{-p^z}{\epsilon + m}
    ),
  \label{eq:spinors}
\end{equation}
while

\begin{equation}
  \bar{u}^s(p) = u^{s\dag}(p)\gamma^0.
  \label{eq:bar}
\end{equation}
As justified in section \ref{sec:ee}, we need only evaluate the t-channel
contribution to $\mathcal{M}$ and multiply the result by 2.  That is, 

\begin{equation} 
  \label{eq:t} 
  \begin{aligned} 
    \mathcal{M}(p_4p_3\leftarrow
    p_2p_1) 
    &\sim
    e^2 \sum_{s_1,s_2,s_3, s_4}
      \bar{u}^{s_4}\left(p_4\right)\gamma^{\mu}u^{s_1}\left(p_1\right)
      \left(\frac{1}{p_3 - p_2}\right)^2
      \bar{u}^{s_3}\left(p_3\right)\gamma_{\mu}u^{s_2}\left(p_2\right).
  \end{aligned} 
\end{equation}
Substituting equations (\ref{eq:diracBasis}) through (\ref{eq:bar}) into
(\ref{eq:t}) yields $\mathcal{M}$ in terms of the components of the electrons'
4-momenta.  That is,

\begin{equation}
\begin{aligned}
\mathcal{M}(p_4p_3\leftarrow p_2p_1)
&\sim
-\frac{2e^2}{(p_3 - p_2)^2}
\big[(\epsilon_1 + m)(\epsilon_2 + m)(\epsilon_3 + m)(\epsilon_4 + m)\big]^{-1/2}
\\&\times\Big\{
  (\epsilon_1 + m)p_4^x\big[(\epsilon_2 + m)p_3^x
    + (\epsilon_3 + m)p_2^x\big]
  \\&\qquad+
  2(\epsilon_1 + m)(\epsilon_3 + m)(p_2^y + ip_2^z)(p_4^y - ip_4^z)
  \\&\qquad+
  2(\epsilon_2 + m)(\epsilon_4 + m)ip_1^z(p_3^y - ip_3^z)
  \\&\qquad-
  \big[(\epsilon_2 + m)(\epsilon_3 + m) + p_2^xp_3^x
      + (p_2^y + ip_2^z)(p_3^y - ip_3^z)\big]
  \\&\qquad\qquad\times
  \big[(\epsilon_1 + m)(\epsilon_4 + m) + ip_1^z(p_4^y - ip_4^z)\big]
\Big\},
\end{aligned}
\end{equation}
where we let $p_1$ denote the momentum of the beam electron so that $p_1^x =
p_1^y = 0$.
See the Mathematica \cite{Mathematica} notebook in the supplemental material
for more details.

\pagebreak
\section{Normalization}
\label{app:normalization}

When integrating over 4-momentum space as is done in section
\ref{sec:wavepackets}, Lorentz invariance constrains a particle's 4-momentum to
obey $p^2 = m^2$.
It follows that the 4-momentum integration measure $d^4p$ is always multiplied
by a delta function $\delta(p^2-m^2)$, i.e.,

\begin{equation}
    d^4p\delta(p^2 - m^2)\theta(p^0)
    =
    \frac{d^4p}{2p^0}\delta(p_0 - \epsilon_\mathbf{p}),
\end{equation}
where the Heaviside step function restricts our consideration to particles of
positive mass (we note that antiparticles are interpreted as positive mass
particles that propagate backwards in time).  With this integration measure,
the identity operator can be written as

\begin{equation}
\label{eq:identity}
\begin{aligned}
    \hat{I}
    &=
    \int \frac{d^4p}{(2\pi)^4}(2\pi)\delta(p^2 - m^2)\theta(p^0)
    \ket{p}\bra{p}
    \\&=
    \int \frac{d^3p}{(2\pi)^3 2\epsilon_\mathbf{p}}
    \ket{p}\bra{p}
    \\&=
    \int \frac{d^3p}{(2\pi)^3}
    \ket{\mathbf{p}}\bra{\mathbf{p}}.
\end{aligned}
\end{equation}
The last equality implies that

\begin{equation}
\label{eq:normalization}
    \ket{p} = (2\epsilon_\mathbf{p})^{1/2}\ket{\mathbf{p}}.
\end{equation}

\pagebreak
\section{Lab frame $p^z_3$}
\label{app:p3z}

For the scattering of two free electrons, conservation of 4-momentum allows us
to write

\begin{equation}
    p_1 + p_2 = p_3 + p_4.
\end{equation}
This constrains four of the six components needed to specify the 3-momenta of the
two outgoing particles, so that only two componenets are independent.
We choose the independent components to be $p^x_3$ and $p^y_3$.  Thus, we wish
to find $p^z_3$ as a function of $p_1$, $p_2$, $p^x_3$, and $p^y_3$.
From this, $p_4 = p_1 + p_2 - p_3$ is easily obtained, and we have all six
components of outgoing 3-momenta needed to calculate the scattering amplitude
from equation (\ref{eq:ampCode}).
We start by setting the $z$-direction parallel to $\mathbf{p}_1$.
This means that

\begin{equation}
\label{eq:conservation}
    \begin{aligned}
        &p^x_4 = p^x_2 - p^x_3
        \\&p^y_4 = p^y_2 - p^y_3
        \\&p^z_4 = p^z_1 + p^z_2 - p^z_3
        \\&\epsilon_4 = \epsilon_1 + \Delta\epsilon
    \end{aligned}
\end{equation}
where

\begin{equation}
  \epsilon_i = \epsilon_{\mathbf{p}_i} = \sqrt{\mathbf{p}_i^2 + m^2},
  \qquad\quad
  \Delta\epsilon = \epsilon_{n\mathbf{k}} - \epsilon_{n'\mathbf{k'}},
\end{equation}
and $\epsilon_{n\mathbf{k}}$ and $\epsilon_{n'\mathbf{k'}}$ are the eigenvalues
of the excited hole and electron states, respectively.
We can also write $\epsilon_1 = \gamma m$ where $\gamma=1/\sqrt{1-\beta^2}$ is
the Lorentz factor corresponding to the beam electron's velocity $\beta$.
Squaring the last line in (\ref{eq:conservation}) and subtracting $m^2$ then
gives

\begin{equation}
  \mathbf{p}_4^2
  =
  \mathbf{p}_1^2 + 2\gamma m\Delta\epsilon + \mathcal{O}(\Delta\epsilon^2).
  \label{eq:p41}
\end{equation}
We can ignore terms of order $\mathcal{O}(\Delta\epsilon^2)$ since
$\Delta\epsilon\ll m$.  Meanwhile, squaring and summing the first three
equations in (\ref{eq:conservation}) tells us that

\begin{equation}
  \mathbf{p}_4^2
  =
  \mathbf{p}_1^2 + \mathbf{p}_2^2 + \mathbf{p}_3^2
  + 2\left[
  p^z_1p^z_2 - \mathbf{p}^{\perp}_2 \cdot \mathbf{p}^{\perp}_3 
    - (p^z_1 + p^z_2)p^z_3
  \right],
  \label{eq:p42}
\end{equation}
where the $\perp$ superscript denotes the projection perpendicular to
$\hat{z}$.
Subtracting equation (\ref{eq:p42}) from equation (\ref{eq:p41}) then yields

\begin{equation}
\label{eq:eliminatep4}
\begin{aligned}
    0
    &\sim
    \mathbf{p}_2^2 + \mathbf{p}_3^2
    +
    2\left[
    p^z_1p^z_2 - \mathbf{p}^{\perp}_2 \cdot \mathbf{p}^{\perp}_3 - (p^z_1 + p^z_2)p^z_3
    \right]
    -
    2\gamma m\Delta\epsilon.
\end{aligned}
\end{equation}
Asymptotic formula (\ref{eq:eliminatep4}) can then be solved for $p^3_z$, so that

\begin{equation}
\label{eq:p3z}
    p_3^z
    \sim
    p_1^z + p_2^z
    \pm
    \sqrt{
    \left(p_1^z\right)^2
    -
    \left(\mathbf{p}_2^\perp - \mathbf{p}_3^\perp\right)^2
    +
    2\gamma m\Delta\epsilon
    }.
\end{equation}
We choose the $-$ from $\pm$ as we impose that $\mathbf{p}_3$ is a
component of the outgoing crystal electron state, whose $z$-momentum is much
smaller than that of the beam electron.

\pagebreak
\section{Converging $n_i^\text{max}$}
\label{app:nimax}

\begin{figure}[H]
  \centering
  \includegraphics[width=\textwidth]{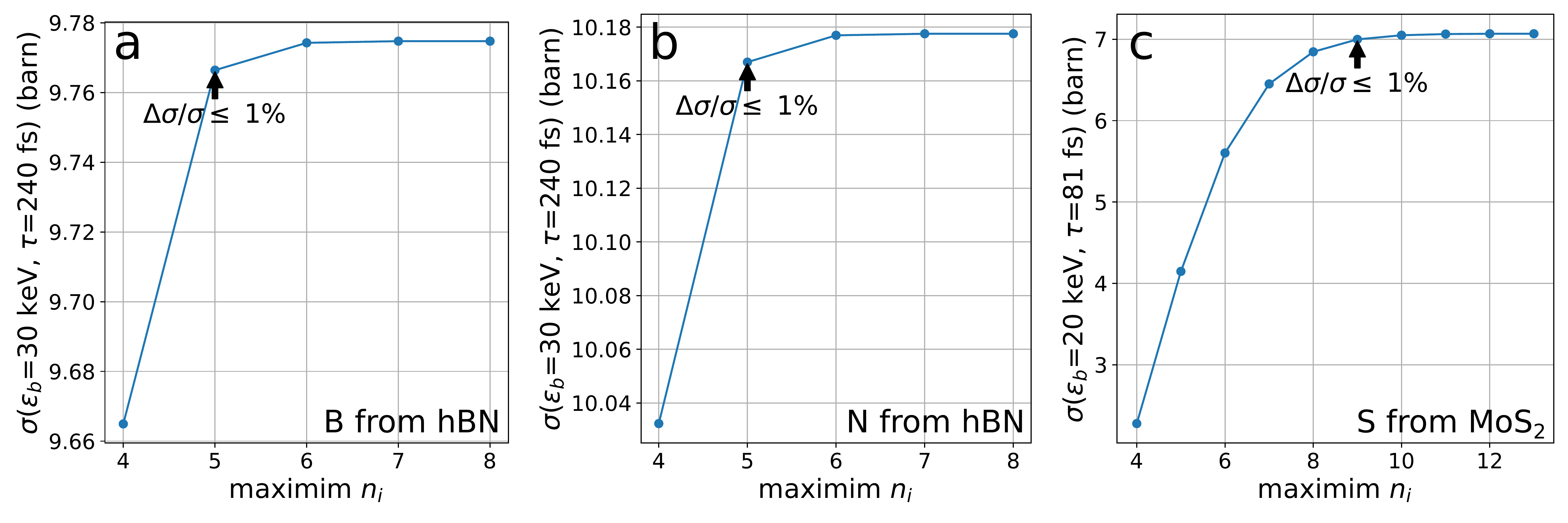}
  \caption{
    Convergence of the sputtering cross section with respect to the maximum
    number of beam-induced excitations $n_i^\text{max}$ considered.
    The simulated beam energies $\epsilon_b$ are the lowest experimental beam
    energies used for each material.
    The excitation lifetimes $\tau$ are those used to fit the experimental data
    in figures \ref{fig:edgeCross} and \ref{fig:MoS2Cross}.
    Each cross section is deemed converged when any increase in $n_i^\text{max}$
    increases the cross section by less than 1\%.
  }
  \label{fig:nimax}
\end{figure}

\pagebreak
\section{Peaks in the sputtering cross section of hBN}
\label{app:edgePeaks}

\begin{figure}[H]
  \centering
  \includegraphics[width=0.6\textwidth]{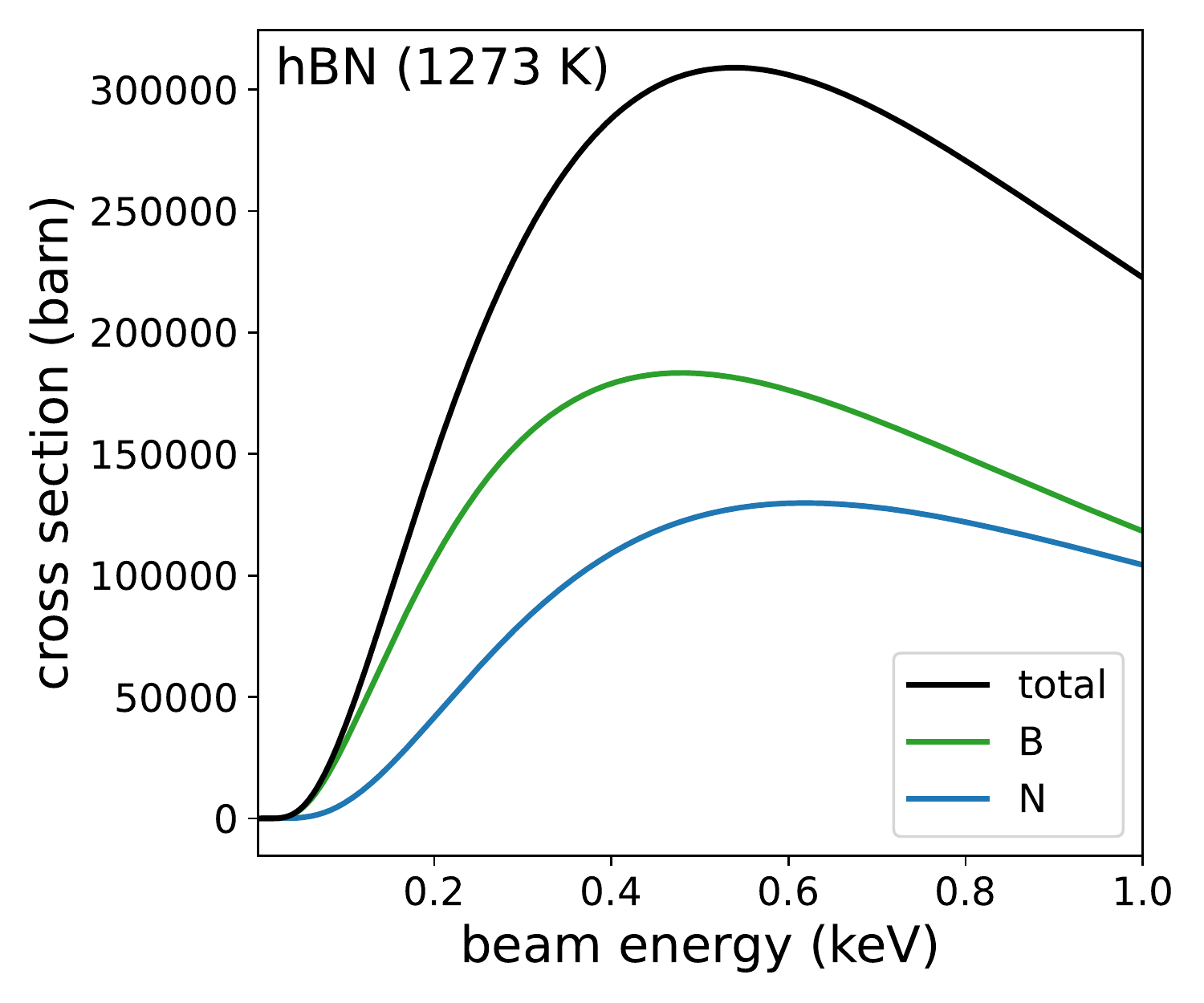}
  \caption{
    The sputtering cross sections of boron and nitrogen from the hBN armchair
    edge peak at beam energies much lower than those typically used for
    microscopy and defect engineering.
    However, as mentioned in the main text, the validity of our perturbative
    approximation to the scattering operator breaks down at low beam energies
    (section \ref{sec:Pi}).
    Thus, the values and positions of these peaks may change if higher order
    perturbation terms are considered.
  }
  \label{fig:edgePeaks}
\end{figure}

\pagebreak
\section{Fitting and converging $S$}
\label{app:fitting}

\begin{figure}[H]
  \centering
  \includegraphics[width=\textwidth]{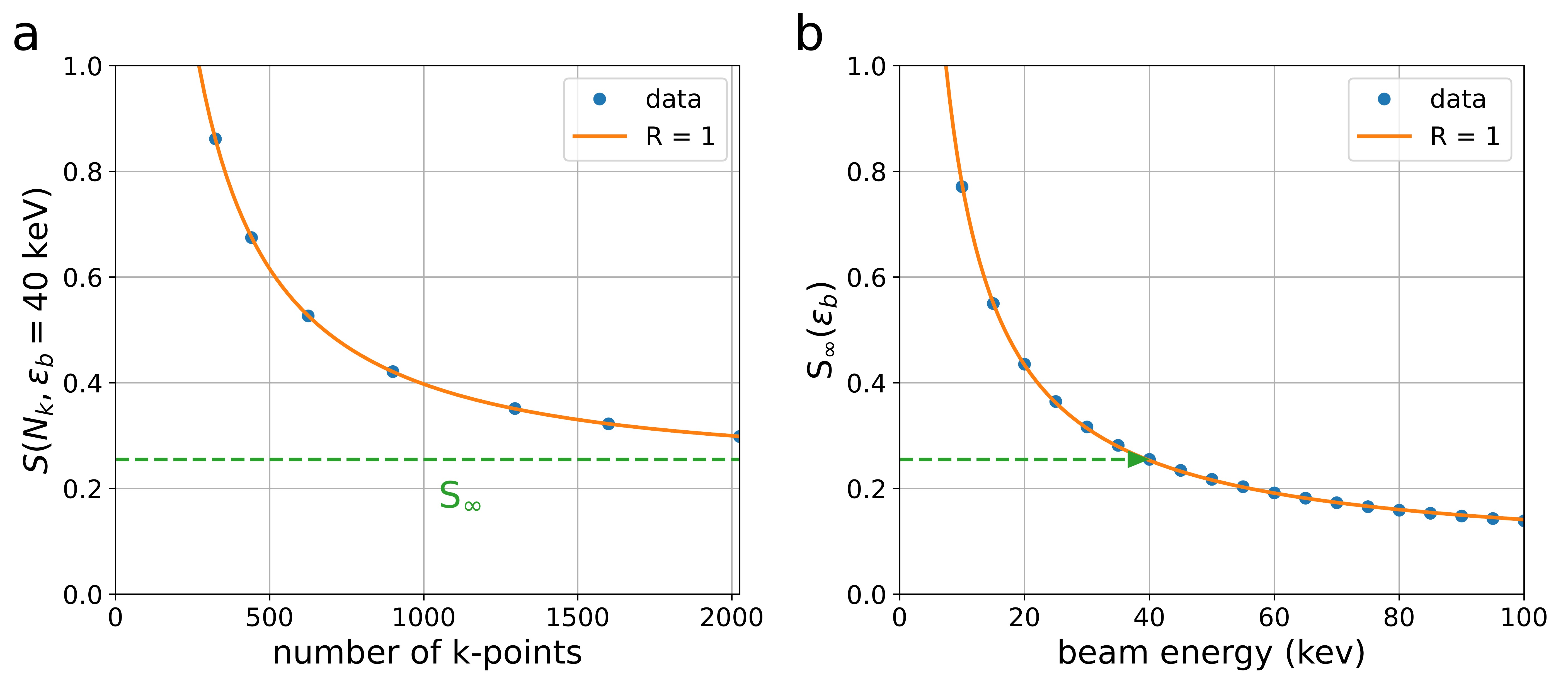}
  \caption{
    Finding the large crystal limit of $S(\epsilon_b)$ requires extrapolation.
    Panel (a) shows the dependence of $S$ on the number k-points $N_k$ in the
    Brillouin zone of hBN under 40 keV irradiation.
    The simulated points are fitted well by equation (\ref{eq:SNk}).
    The green dashed line denotes $S_\infty$, the asymptotic limit of $S(N_k)$
    for large $N_k$.
    Panel (b) then shows the dependence of $S_\infty$ on the beam energy
    $\epsilon_b$, which is fitted well by equation (\ref{eq:SEb}).
    The green arrow in panel (b) illustrates how $S_\infty(\epsilon_b=40\text{
    keV})$ is determined by the fit in panel (a).
  }
  \label{fig:Sfit}
\end{figure}

For a given beam energy $\epsilon_b$, we calculate the sum of all transition
probabilities $S$ defined in equation (\ref{eq:S}) for various k-point
densities where the number of k-points in the Brillouin zone is $N_k$.
We then fit the points to a curve of the form

\begin{equation}
  S(N_k, \epsilon_b)
  =
  \frac{a}{N_k-b}e^{-cN_k} + S_\infty,
  \label{eq:SNk}
\end{equation}
where $a$, $b$, $c$, and $S_\infty$ are all fitting parameters that depend on
$\epsilon_b$ (figure \ref{fig:Sfit}a).
We repeat this for multiple values of $\epsilon_b$ ranging from 5 to 100 keV
and record the best fit
$S_\infty$ for each
$\epsilon_b$.
Based on the work of Bethe \cite{Bethe1930, Susi2019, Kretschmer2020}, we fit the resulting values of
$S_\infty(\epsilon_b)$ to an inverse function,

\begin{equation}
  S_\infty(\epsilon_b)
  =
  \frac{A}{\epsilon_b - B} + C.
  \label{eq:SEb}
\end{equation}
where $A$, $B$, and $C$ are fitting parameters.  The parameters for hBN and
MoS$_2$ are given in table \ref{tab:fit}.
The fitted curve can then be substituted for $S(\epsilon_b)$ in formula
(\ref{eq:Pi}) to obtain $P_i(\epsilon_b, n_i)$, the probability of exciting
$n_i$ electrons in the large crystal limit.

\begin{table} \centering 
  \begin{tabular}{lccc}
    \toprule
    material &A (eV) &B (eV) &C \\
    \midrule
    hBN     &7.655 &-0.770 &0.06526 \\
    MoS$_2$ &49.05 &-3.867 &0.3547 \\
    \bottomrule
  \end{tabular}
  \caption{
    Fitting parameters of equation (\ref{eq:SEb}) for hBN and MoS$_2$.
  } 
\label{tab:fit}
\end{table}

\begin{figure}[H]
  \centering
  \includegraphics[width=\textwidth]{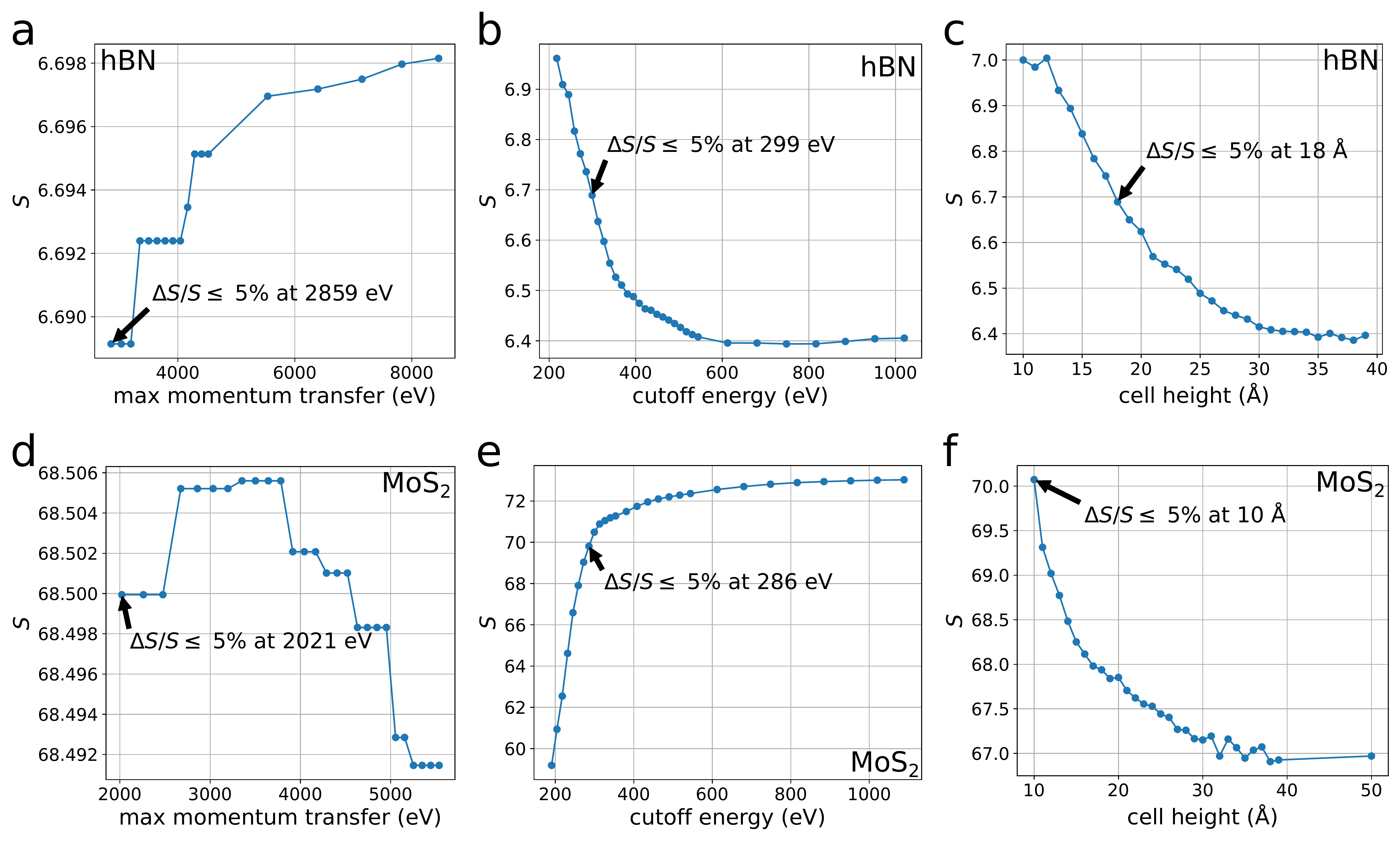}
  \caption{
    Convergence of $S$ with respect to the (a and d) maximum magnitude of
    virtual photon momentum considered, (b and e) plane-wave DFT cutoff energy,
    and (c and f) height of the unit cell.
    A $6\times6\times1$ k-point mesh and a beam energy of 60 keV is used to
    generate all six plots.
    The converged parameters were found to be insensitive to changes in k-point
    density and beam energy.
    $S$ is deemed converged when any increase in precision changes $S$ by less
    than 5\%.
  }
  \label{fig:convergences}
\end{figure}

Finally, $S$ must also be converged with respect all parameters.  These include
the maximum virtual photon momentum, DFT cutoff energy, and height of the
pristine unit cell (figure \ref{fig:convergences}).
$S$ is considered converged with respect to a parameter when any increase in
the parameter's precision changes $S$ by less than 5\%.

\break

\pagebreak
\section{Calculating $E_\text{min}$}
\label{app:Emin}

\begin{figure}[H]
  \centering
  \includegraphics[width=0.6\textwidth]{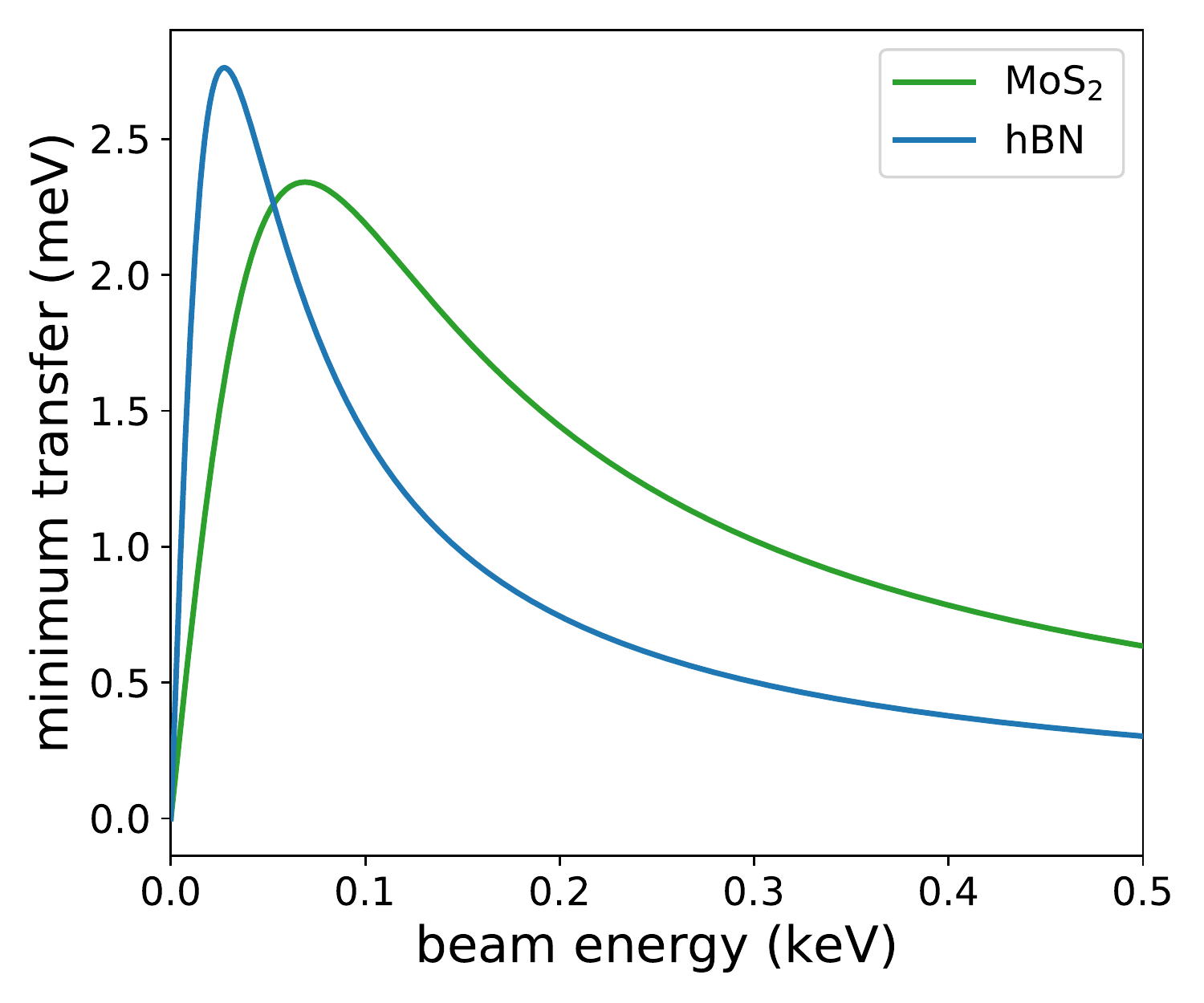}
  \caption{
    The minimum energy transfer from the beam electron to the target nucleus
    peaks at very low beam energies and is always much smaller than the
    nuclei's average thermal kinetic energy of $\sim$39 meV at room
    temperature. 
  }
  \label{fig:Emin}
\end{figure}

In hBN, a unit cell contains a boron and nitrogen atom.
As these atoms have similar masses and displacement thresholds (table
\ref{tab:Ed}), the sputtering of both atoms should be considered for a given
beam energy.
Thus, we approximate the maximum cross sectional area $\sigma_\text{max}$ of
these atoms to be $\Omega_\text{hBN}/2$, where $\Omega_\text{hBN}$ is the area
of the hBN unit cell.
On the other hand, molybdenum is much heavier than sulfur and has a much larger
displacement threshold in MoS$_2$ \cite{Komsa2012}.  
This means that only the sputtering of sulfur needs to be considered.
Additionally, of the two sulfur atoms in the MoS$_2$ unit cell, only the atom
on the outgoing surface is eligible to sputter from a pristine system
\cite{Komsa2012}.
Therefore, $\sigma_\text{max}$ for sulfur sputtering from MoS$_2$ is
$\Omega_\text{MoS$_2$}$, the area of the MoS$_2$ unit cell.

To approximate $E_\text{min}$, we use the Rutherford displacement cross section
\cite{Thornton2004, Sakurai2011, Yoshimura2018} as an approximation to equation
(\ref{eq:basicSigma}),

\begin{equation}
  \sigma_R
  =
  \pi\left(\frac{Z\alpha}{|\mathbf{p}|\beta}\right)^2
  \left(\frac{E_\text{max}}{E_d} - 1\right).
  \label{eq:Rutherford}
\end{equation}
Setting $\sigma_R=\sigma_\text{max}$ and $E_d=E_\text{min}$ and solving for
$E_\text{min}$ yields

\begin{equation}
  E_\text{min}(\epsilon_b)
  =
  E_\text{max}
  \left[\frac{\Omega}{\pi}
    \left(\frac{|\mathbf{p}_b|\beta}{Z\alpha}\right)^2 + 1
  \right]^{-1}.
  \label{eq:Emin}
\end{equation}
Using the Rutherford cross section instead of the McKinley-Feshbach
cross section should be accurate for small beam energies for which $\beta\ll
1$.
This makes it well-suited for finding $E_\text{min}$, since
$E_d(n_f)<E_\text{min}$ only for large $n_f$, and large $n_i$ (and thus $n_f$)
are much more likely for small beam energies (figure \ref{fig:Pi}).

With that said, the sputtering cross section is fairly insensitive to the exact
value of $E_\text{min}$ when $E_\text{min} \ll E_\text{max}$.
This is because the post-collision velocity of the PKA diminishes when the
energy transfer $E$ shrinks.
The beam-induced excitations therefore have more time to relax before the PKA
reaches the step in the energy surface (section \ref{sec:assumptions}).
This makes $P_f$ small for small $E$, so that contributions to the integral in
equation (\ref{eq:totSigma}) are extremely tiny for $E$ near $E_\text{min}$.
As a result, changes in $E_\text{min}$ are essentially immeasurable for beam
energies greater than 1 keV.
Nonetheless, the use of $E_\text{min}$ is necessary for the calculation of a
finite sputtering cross section.

\makeatletter\@input{saux.tex}\makeatother

\bibliography{library}
\bibliographystyle{ieeetr}